\documentclass[reprint,aps,twoside,prd,superscriptaddress,nofootinbib,longbibliography]{revtex4-2}

\pdfoutput=1 
\usepackage[T1]{fontenc} 
\usepackage{dcolumn}
\usepackage{physics} 
\usepackage{makecell}
\usepackage{enumerate}

\usepackage{amssymb}
\usepackage{amsmath}
\usepackage{amsthm}
\usepackage{tensor,mathtools}
\usepackage{nicefrac}
\usepackage{cancel}

\usepackage[english]{babel}
\usepackage{amsthm}

\theoremstyle{definition}
\newtheorem*{definition}{Definition}

\usepackage[version=4]{mhchem}
\usepackage[dvipsnames]{xcolor}
\usepackage{graphicx}
\usepackage{url}
\usepackage[colorlinks]{hyperref}

\hypersetup{
     breaklinks=true,
    pdfstartview={FitH},    
    colorlinks=true,       
    linkcolor=blue,          
    citecolor=red,        
    filecolor=magenta,      
    urlcolor=blue,           
    anchorcolor=green,      
    linktocpage=true
}

\date{\today}

\begin{document}

\title{Locally Rotationally Symmetric Spacetimes in Einstein-Cartan Theory and Their Classification}

\author{Ujjwal Agarwal}
\email{ujjwal.agarwal@matfyz.cuni.cz}
\affiliation{Institute of Theoretical Physics, Faculty of Mathematics and Physics,
Charles University, Prague, V Hole{\v s}ovi{\v c}k{\' a}ch 2, 180 00 Prague 8, Czech Republic}

\author{Sante Carloni}
\email{sante.carloni@matfyz.cuni.cz}
\affiliation{Institute of Theoretical Physics, Faculty of Mathematics and Physics,
Charles University, Prague, V Hole{\v s}ovi{\v c}k{\' a}ch 2, 180 00 Prague 8, Czech Republic}
\affiliation{DIME Sez. Metodi e Modelli Matematici, Universit\`{a} di Genova, Via  All’Opera Pia 15, 16145 - Genoa, (Italy).}
\affiliation{INFN Sezione di Genova, Via Dodecaneso 33, 16146 Genova, Italy}

\begin{abstract}
We present the complete set of covariant equations that govern the locally rotationally symmetric torsion spacetimes sourced by Weyssenhoff fluid in Einstein-Cartan-Sciama-Kibble gravity.  Using these equations, we can explore in detail the peculiar relationship between conformal structure and torsion. We develop a comprehensive scheme to categorize these torsional spacetimes into distinct classes. We explicitly analyze the properties of each class and obtain novel analytical solutions to the gravitational field equations.
\end{abstract}
\maketitle

\section{Introduction}

The issue of extending the classical scheme of General Relativity (GR) is a widely discussed subject of research. While GR can be extended in many different ways, one of the most studied extensions is achieved by relaxing the condition of torsion-free covariant derivative. The reason for such interest in the inclusion of torsion goes beyond the theoretical exploration of the mathematical structure of pseudo-Riemannian spacetimes. 

Historically, General Relativity was not designed to account for quantum effects, such as spin. The inclusion of spin and the study of its influence on the differential structure of a manifold, thus, is a problem of fundamental importance to understand the interaction between gravity and the microphysical domain in which quantum phenomena play a significant role. While there is no unique prescription to include spin to the differential structure of the manifold, one of the most widely studied theories that achieves this integration is the Einstein-Cartan-Sciama-Kibble (ECSK) theory of gravity (sometimes referred to as Einstein-Cartan theory) \citep{Hehl}. 

ECSK theory is a consistent extension of GR which postulates that the intrinsic spin of matter fields is coupled to a non-vanishing torsion tensor field \citep{Hehl,Sciama,Kibble}. The reason behind such a choice is that spin is related to the rotational part of the Poincar\'{e} group, whereas mass is connected to its translational part. Thus, the gravitation of spin must be represented separately from the one of mass. This can be achieved by introducing \textit{Hypermomentum tensor} (or spin \textit{angular momentum tensor})  on top of the standard energy-momentum tensor to describe the effects of spin on spacetime.  In the field equations, the hypermomentum tensor couples to a geometrical quantity, which is a combination of the torsion tensor, associated with the rotational degrees of freedom in spacetime \citep{Hehl}.

Interestingly, in ECSK theory, the coupling of spin to torsion leads to certain modifications in the field equations that are quadratic in spin terms and do not vanish on spacetime averaging (despite the dipole character of spin) \citep{Hehl,HehlSingularity}. This fact highlights that even in the macroscopic domain, the ECSK theory predicts non-vanishing effects of spin on the spacetime geometry. In particular, these effects can be measurable at extremely high densities of matter \citep{HehlSingularity}. 

At macroscopic scale, one can describe the matter fluid with spin via the Weyssenhoff fluid \citep{HehlSingularity,WeyssenhoffRaabe,Korotky,Halbwachs}. The Weyssenhoff fluid is a semi-classical model of a spin matter fluid, which at macroscopic scales, following an integration over small volume elements, reduces to the Frenkel-Mathisson equations \citep{Mathisson1937,Frenkel,Paparetrou}.

One type of macroscopic system where both spin and gravitation play a significant role is compact stars, in particular, neutron stars.
A neutron star is a collapsed core of a massive star that is formed following a supernova explosion. Its core, composed of fermionic matter at extreme density in a strong gravity environment, is made stable by the neutron degeneracy pressure
\citep{PotekhinPhysicsofNS}. Thus, a neutron star serves as an ideal laboratory in which theoretical models, such as the ECSK theory of gravity, can be tested, particularly against the recently available gravitational wave data \citep{Abbott_2017}.

While the above discussion focuses on the inclusion of spin matter fields, torsion also provides an alternative explanation for several open problems in gravitational physics. For example, in cosmology, the introduction of torsion can offer theoretical frameworks for the late-time acceleration of the universe, introduce an inflationary period at very early times, and avoid the Big Bang singularity \citep{CaiCapozziello-f(T)gravity,Kopczynski-Singularity}. Torsion might even prevent complete gravitational collapse to a singularity \citep{Trautman-Singularity,Stewart-Singularity,HehlSingularity}. Various other fields of research in which the inclusion of torsion can be useful are discussed in \citep{Capozziello, CaiCapozziello-f(T)gravity} and references therein.

In summary, the study of spacetimes endowed with torsion could provide a better understanding of the interaction between matter and gravitational field (as in, geometry of the spacetime) in both the macroscopic and microscopic domains. In particular, the inclusion of torsion in the study of cosmological models, the interior solutions of stellar objects, and perturbations of such spacetimes can be quite fruitful. 
However, the task of resolving the gravitational field equations can be very complex in general, forcing one to exploit the symmetries of the spacetimes as much as possible. Luckily, many of the most studied cosmological and astrophysical spacetimes fall in the so-called Locally Rotationally Symmetric (LRS) \citep{Ellis1967,EllisStewart1968,EllisElst} class of spacetimes.

LRS spacetimes are defined as pseudo-Riemannian manifolds that possess, at each point, a continuous isotropy group related to rotations around a local axis of symmetry (hence the name ``locally rotationally symmetric'') \cite{Ellis1967,EllisStewart1968}. In General Relativity, LRS spacetimes sourced by a perfect fluid can be separated into three independent classes \citep{EllisStewart1968,EllisElst}. Such a classification provides two benefits. Firstly, it simplifies the geometry of the spacetime and resolution of the governing equations because many physical quantities, and sometimes even certain derivatives, vanish for each class. Secondly, it enables the development of a taxonomy of exact solutions to the gravitational field equations. In GR, the classification helped further the understanding of many physically relevant cosmological and astrophysical spacetimes, such as the Friedmann-Lemaître-Robertson-Walker models, various Bianchi models, and the interior solutions of stellar objects \citep{EllisElst}.

Among the various approaches for analyzing LRS spacetimes, covariant formalism provides a highly effective methodology \citep{EllisElst1998CM, Clarkson_2003, ClarksonBetschart_2004, Clarkson_2007}. In the covariant formalism, one decomposes the tensors on the manifold by projecting them along timelike or spacelike congruences and on the (locally) orthogonal hypersurface. This allows one to define physically meaningful variables that describe all the properties of a manifold while avoiding confinement to a specific coordinate system. In particular, the covariant formalism has led to the development of powerful methods to study the interior of stellar objects \citep{CarloniTOVIso1Fluid,CarloniTOVAniso1Fluid}.

In this paper, utilizing the covariant formalism, we study the LRS spacetimes endowed with non-vanishing torsion\footnote{Since, typically, LRS spacetimes are considered as a class of spacetimes within General Relativity, we wish to clarify further that we aim to study spacetimes endowed with non-vanishing torsion and which possess a continuous isotropy at each point which is given by a spacelike vector describing a local axis of symmetry.}. We shall refer to these spacetimes as \textit{Torsional Locally Rotationally Symmetric} (TLRS) spacetimes, in analogy with the class of LRS spacetimes in General Relativity \citep{EllisStewart1968,EllisElst}. 
This will enable us to develop a framework that can be utilized to study many physically interesting torsional spacetimes, including cosmological and astrophysical ones. 
We focus mainly on the Einstein-Cartan theory of gravity \citep{Hehl,Kibble,Sciama} sourced by Weyssenhoff fluid \citep{HehlSingularity,WeyssenhoffRaabe,Korotky,Halbwachs}. Thus, as discussed, the torsion would be coupled to the spin of a semi-classical spin matter (fermionic) fluid.
Following 1+1+2 decomposition, we derive the complete set of governing equations for Torsional Locally Rotationally Symmetric spacetimes.
Further, despite the spin matter fluid generating torsion being non-classical and non-perfect, we were able to develop a complete classification scheme and explore some interesting examples that have no corrective in GR.

This paper is organized as follows: in Section \ref{Sec:TorsionSpacetimes}, we define the fundamental quantities that describe the properties of the manifold, such as the torsion and Riemann tensors and their properties. We provide a brief review of Einstein-Cartan gravity and its field equations, as well as the spin matter source of torsion, the Weyssenhoff fluid. 
In Section \ref{Sec:1+1+2Decomposition}, we provide a summary of the 1+1+2 decomposition and define the covariant variables which describe the Torsional Locally Rotationally Symmetric spacetimes.
In Section \ref{TLRSEqnDerivation}, we provide a pedagogical derivation of the covariant equations governing the spacetime, along with the consistency conditions that must be satisfied. We also discuss the condition so that the manifold admits a foliation into a submanifold and a hypersurface orthogonal vector field. We then combine the results to simplify the governing equations and provide a complete set of covariant equations which govern the spacetime.
In Section \ref{Classification-of-LRSGR-Subsection}, we review the classification of LRS spacetimes in General Relativity and highlight some significant features.
In Section \ref{Classification-of-TLRSSpacetimes-subsection}, we describe the classification of the TLRS spacetimes with Weyssenhoff fluid matter source and show that the key features of classification in General Relativity are preserved.
In Section \ref{Sec:ClassIB}, we present some examples of TLRS spacetimes.
Finally, in Section \ref{DiscussConclusion}, we present the conclusions and discuss the results of the paper.

In this article, we use metric signature ${(-+++)}$ and natural units ${8\pi G = c = 1}$. We utilize abstract indices to denote the rank of tensor fields, and help describe their symmetries, and perform operations \citep{WaldBook}. As such, these indices do not denote the components in a specific basis.

\section{Torsion Spacetimes and Einstein-Cartan Theory} \label{Sec:TorsionSpacetimes}

We first define the important quantities needed to describe a $4-$dimensional pseudo-Riemannian manifold $\mathcal{M}$ equipped with a metric tensor $g_{ab}$ and a covariant derivative ($\nabla$), which is metric compatible ($\nabla g = 0$)\citep{Hehl,HawkingEllisBook,WaldBook}. The torsion tensor $T^k{}_{ab}$ is described as
\begin{align}
    \nabla_a\nabla_b \psi - \nabla_b \nabla_a \psi &= -T^k{}_{ab} \nabla_k \psi
    \ , \label{DefofTorsion}
\end{align}
where $\psi$ is a generic function on the manifold. The Riemann tensor $R_{abcd}$ is described as
\begin{align}
    \nabla_a \nabla_b X_c - \nabla_b \nabla_a X_c + T^k{}_{ab} \nabla_k X_c &= -R_{ab}{}^d{}_c X_d \ ,
\end{align}
where $X^a$ is a generic vector field on the manifold. The Ricci tensor $R_{ab}$ and the Ricci scalar $R$ are given as
\begin{align}
    R_{ab} &= g^{mn}R_{manb} \qquad R = g^{ab} R_{ab}
    \ ,
\end{align}
and finally the Weyl tensor $C_{abcd}$ is defined through the decomposition of the Riemann tensor
\begin{align}
    R_{abcd} 
    &= C_{abcd} + \frac{1}{2}\left( R_{ac}g_{bd} + R_{bd}g_{ac} - R_{ad}g_{bc} - R_{bc}g_{ad} \right) 
    \nonumber \\ & \quad
    - \frac{R}{6}\left( g_{ac}g_{bd} - g_{ad}g_{bc} \right)
     \ . \label{DefWeylTensor}
\end{align}
The symmetries of the Riemann tensor are
\begin{align}
    R_{abcd} &= -R_{bacd} = -R_{abdc}
    \ ,
\end{align}
and the Weyl tensor has the same symmetries as the Riemann tensor. Additionally, by definition, the Weyl tensor is completely traceless.

Finally, in this paper, the Bianchi identities 
\begin{align}
    R_{[ab}{}^{n}{}_{c]} &= \nabla_{[a}T^n{}_{bc]} - T^k{}_{[ab} T^n{}_{c]k}
    \ , \label{BianchiIdType1}
    \\
    \nabla_{[a}R_{bc]}{}^{k}{}_{l} &= T^n{}_{[ab}R_{c]n}{}^{k}{}_{l}
    \ , \label{BianchiIdType2}
\end{align}
 will be referred to as \textit{type-I} and \textit{type-II} respectively. If the Riemann tensor is replaced in \eqref{BianchiIdType2} using \eqref{DefWeylTensor}, we obtain, after contracting the indices $a$ and $l$, the following equation for the derivative of the Weyl tensor
\begin{align}
\nabla^d C_{abcd} 
    &= - \nabla_{[a} R_{b]c} - \frac{1}{6} g_{c[a} \nabla_{b]} R - \frac{1}{2} T_{nmk} R_{[a}{}^{nmk} g_{b]c}
    \nonumber \\ & \quad
    - T_{nm[a} \left( g_{b]c} R^{nm} + 2 R_{b]}{}^{nm}{}_{c} \right) - T^n{}_{ab} R_{nc}
    \label{oncecontractedBianchiId} \ .
\end{align}
Further, a contraction of \eqref{oncecontractedBianchiId}, while utilizing $\nabla g = 0$ and the properties of the Weyl tensor to eliminate the LHS, gives the following
\begin{align}
    \nabla^b R_{ab} -\frac{1}{2}\nabla_a R &= - R^{nm}T_{nma} - \frac{1}{2} R_{anmk} T^{nmk}
    \label{twicecontractedBianchiId}
    \ .
\end{align}
Notice that in each case, while deriving \eqref{oncecontractedBianchiId} and \eqref{twicecontractedBianchiId}, only one unique non-vanishing contraction is possible.

\subsection{Einstein-Cartan gravity and Weyssenhoff fluid} \label{ECWFTheory}
Now, we endow the manifold with field equations that relate the Ricci and torsion tensors to the matter content of spacetime. The details of Einstein-Cartan gravity can also be found in Ref. \citep{Hehl,Capozziello,CarloniLuz2019}. However, in this work, we have used different conventions compared to these references.
In Einstein-Cartan theory, the field equations are given as 
\begin{align}
    R_{ab} - \frac{1}{2}g_{ab} R &= S_{ab}
    \ , \label{EinsteinEqn} \\
    T^c{}_{ab} + 2\delta^c_{[a}T^d{}_{b]d} &= 2\Delta^c{}_{ab}
    \ , \label{HypermomentumEq}
\end{align}
where $S_{ab}$ is the canonical energy-momentum tensor and $\Delta^c{}_{ab}$ is the hypermomentum tensor, or spin angular momentum tensor. Notice that $\Delta^c{}_{ab}$ can be defined as the variation of the matter Lagrangian $L_m$ with respect to a combination of the torsion tensor $T_c{}^{ab}$ called  contortion  $K_c{}^{ab}$
\begin{align}
    \Delta^c{}_{ab} &= -\frac{\delta {L}_m}{\delta K_c{}^{ab}}=-\Delta^c{}_{ba} \ ,
\end{align}
where contortion tensor is defined as
\begin{align}
    K_{cab} = T_{cab} + T_{bac} - T_{abc} = -K_{cba} \ .
\end{align}
However, the same is not true for the canonical energy-momentum tensor $S_{ab}$. Indeed, the variation of the matter Lagrangian  with respect to the metric gives the metric energy-momentum tensor $s_{ab}$
\begin{align}
    s_{ab} &= \frac{\delta {L}_m}{\delta g^{ab}} \ .
\end{align}
which is symmetric by definition (while $S_{ab}$ has no symmetries), and it is related to the canonical energy-momentum tensor by
\begin{align}
    s^{ab} &= S^{ab} + (\nabla_c + T^m{}_{cm}) (\Delta^{abc} - \Delta^{cab} - \Delta^{bca}) \ . \label{MetCanonEMRelBad}
\end{align}
Contracting the Bianchi identity type-I \eqref{BianchiIdType1} and using \eqref{EinsteinEqn} and \eqref{HypermomentumEq}, we obtain
\begin{align}
S_{[ab]} = (\nabla_c + T^m{}_{cm})\Delta^c{}_{ab} \ . \label{AntiSymmCanonEMTnsr}
\end{align}
Equation \eqref{AntiSymmCanonEMTnsr} allows us to write \eqref{MetCanonEMRelBad} in explicitly symmetrical form as
\begin{align}
    s^{ab} &= S^{(ab)} + 2(\nabla_c + T^m{}_{cm}) \Delta^{(ab)c} \ .
    \label{metriccanonicalEMrelation}
\end{align}
Also, equation \eqref{twicecontractedBianchiId} can now be re-written, using \eqref{EinsteinEqn}, as the conservation equation
\begin{align}
    \nabla_b S^{ab} &= \frac{1}{2} T^{n}{}_{n}{}^{a} g^{km}S_{km} - T^{nma} S_{nm} - \frac{1}{2} T_{nmk} R^{anmk}
    \label{ConservationEqn} \ .
\end{align}

We will consider the matter source to be the uncharged Weyssenhoff fluid. The Weyssenhoff fluid  \citep{WeyssenhoffRaabe,Korotky,CarloniLuz2019} is a fluid which carries an anti-symmetric spin density tensor $L_{ab}$ orthogonal to its fluid 4-velocity $u^a$
\begin{align}
    L_{ab} &= - L_{ba}
    \ , &
    u^a L_{ab} &= 0
    \ .
\end{align}
The spin density tensor $L_{ab}$ is postulated to be related to the hypermomentum tensor $\Delta^c{}_{ab}$ and the canonical energy-momentum tensor $S_{ab}$ as
\begin{align}
    \Delta^c{}_{ab} &= u^c L_{ab}
    \label{spinhypermomWFpostulate}\ , \\
    S_{ab} &= -u_a P_b + p (g_{ab} + u_a u_b)
    \label{EMVctrWFpostulate} \ ,
\end{align}
where  $p$ is the isotropic pressure in the rest frame of the fluid, and $P_a$ is the 4-vector density of energy-momentum. Contracting \eqref{HypermomentumEq} with $\delta^b_c$ and using \eqref{spinhypermomWFpostulate}, one finds that for Weyssenhoff fluid, the compatible torsion tensor must satisfy
\begin{align}
    T^k{}_{ak} = 0 \ . 
\end{align}
This simplifies \eqref{HypermomentumEq} and gives the solution
\begin{align}
    T^c{}_{ab} = 2\Delta^c{}_{ab} = 2 u^c L_{ab} \ . \label{SolofHyperMomeq}
\end{align} 
Finally, we need to resolve the relation between the 4-vector density $P_a$ and the spin density tensor $L_{ab}$. Taking the projection of \eqref{AntiSymmCanonEMTnsr} along $u^a$, using \eqref{EMVctrWFpostulate} to replace $S_{ab}$, and defining ${\mu = u^a u^b S_{ab} = u^a P_a}$ as the energy density in the rest frame of the fluid, we obtain
\begin{align}
    P_a &= -\mu u_a - 2 \Delta^m{}_{na} \nabla_m u^n \ ,
\end{align}
which leads us to the canonical energy-momentum tensor of the Weyssenhoff fluid
\begin{align}
    S_{ab} = \mu u_a u_b + p (g_{ab} + u_a u_b) - 2 u_a L_{bn} u^m \nabla_m u^n
    \ . \label{WeysennFluidEMTnsr}
\end{align}

\section{The 1+1+2 Description of LRS Spacetimes with Torsion} \label{Sec:1+1+2Decomposition}

We provide here a concise summary of the $1+1+2$ covariant approach. The complete description can be found in Refs. \citep{EllisElst1998CM,Clarkson_2003,ClarksonBetschart_2004,Clarkson_2007}.
In the 1+1+2 covariant approach, the tensors and equations on the 4-dimensional manifold are decomposed by considering two congruences, a generic normalized timelike vector field $u^a$ and a normalized spacelike vector field $e^a$
\begin{align}
    u^a u_a &= -1 \ , & e^a e_a &= 1 \ ,
\end{align}
which are orthogonal\footnote{The concept of orthogonality here is different from the notion of hypersurface orthogonality of a congruence to an integral submanifold. We discuss hypersurface orthogonality in detail in Section \ref{FoliationoManifoldConsistencyConditions}.}
\begin{align}
    & u^a e_a=0 \ ,
\end{align}
and the projection operators $h_{ab}$ and $N_{ab}$ are defined as
\begin{align}
    h_{ab} &= g_{ab} + u_a u_b 
    \ , &
    u^a h_{ab} &= 0
    \ , \\
    N_{ab} &= h_{ab} - e_a e_b
    \ , &
    u^a N_{ab} &= 0
    \ , &
    e^a N_{ab} &= 0
    \ .
\end{align}
The projection operator $h_{ab}$ can describe the geometry of the 3-hypersurface orthogonal to $u^a$ locally. Similarly, the projection operator $N_{ab}$ can describe the geometry of a 2-surface orthogonal to $u^a$ and $e^a$ locally. The benefit of such a formalism is that one is not confined to a specific coordinate system. Yet, the 1+1+2 covariant variables described via projections of tensors have a rigorous mathematical definition and a distinct physical meaning.

The Levi-Civita tensor for 4-dimensional spacetime ${\eta_{abcd}= \eta_{[abcd]}}$ is
\begin{equation}
\begin{aligned}
    \eta_{0123} &=  \sqrt{|g|}\ ,&& g = det(g_{ab})
    \ , \\
    \eta^{0123} &= \frac{sign(g)}{\sqrt{|g|}}\ , &&  sign(g) = g/|g|
    \ .
\end{aligned}
\end{equation}
The projections of the Levi-Civita tensor are defined as
\begin{align}
    \eta_{abc} &= \eta_{[abc]} = \eta_{dabc} u^d
    \ , &
    \eta_{ab} &= \eta_{[ab]} = \eta_{abc} e^c
    \ ,
\end{align}
which can be related to each other as
\begin{align}
    \eta_{abc} &= e_a \eta_{bc} - e_b \eta_{ac} + e_c \eta_{ab} \ .
\end{align}

In general, the contraction of Levi-Civita tensor for a \textit{d}-dimensional manifold described by metric tensor $P_{ab}$ is given as
\begin{align}
    \eta_{a_1 ... a_m c_1 ... c_n} \eta^{b_1 ... b_m c_1 ... c_n} &= sign(P) \ m! \ n! \ P^{[b_1}_{a_1} ... P^{b_m ]}_{a_m} 
\end{align}
where we have ${m+n=d}$, ${P = det(P_{ab}(x))}$  and ${sign(P) = P/|P|}$. Specifically, in the case of the $2-$dimensional Levi-Civita tensor, we have
\begin{align}
    \eta^{ab}\eta_{pq} &= N^a_p N^b_q - N^a_q N^b_p
    \ , &
    \eta^{ab}\eta_{pb} &= N^a_{p}
    \ .
\end{align}
For convenience, we define some shorthand notations. The covariant derivative along the timelike vector field $u^a$ is written as the \textit{dot derivative}
\begin{align}
    u^a \nabla_a T^b{}_c &= \dot{T}^b{}_c \ ,
\end{align}
and full spatially projected covariant derivative is written as $(\tilde{\nabla}_a)$
\begin{align}
    h_a^p h^b_q h_c^r \nabla_p T^q{}_r &= \tilde{\nabla}_a T^b{}_c \ ,
\end{align}
which can then be projected to define the covariant derivative along the vector field $e^a$ given as the  \textit{hat derivative}
\begin{align}
    e^a \tilde{\nabla}_a T^b{}_c &= \hat{T}^b{}_c \ ,
\end{align}
where the tensor $T^a{}_b$ is a generic tensor on the manifold. 
The full projection of the covariant derivative onto the 2-surface described by $N_{ab}$ is written as ($\overline{\nabla}_a$)
\begin{align}
 N_a^p N^b_q N_c^r \tilde{\nabla}_p T^q{}_r 
    &= \overline{\nabla}_a T^b{}_c 
    \ ,
\end{align}
where the tensor $T^a{}_b$ is a generic tensor on the manifold.

In this paper, we consider spacetimes which are locally rotationally symmetric (LRS) \citep{EllisStewart1968,EllisElst}. This implies that the spacetime possesses a local axis of symmetry at each point on the spacelike hypersurface described by $h_{ab}$. This local axis of symmetry can be described as a spacelike vector field.  We choose the congruence $e^a$ to be parallel to this preferred direction. Thus, all vector and tensor fields orthogonal to $u^a$ and $e^a$, that is, on the surface $N_{ab}$, vanish, making the 1+1+2 covariant formalism particularly well suited to study LRS spacetimes.  As mentioned in the introduction, and in line with the literature, we assign a new name to torsional spacetimes that possess local rotational symmetry: \textit{Torsional Locally Rotationally Symmetric} or {\it TLRS } spacetimes.

The key 1+1+2 variables that  describe TLRS spacetimes can be obtained by projecting and decomposing the covariant derivatives of $u^a$ and $e^a$ and the Weyl tensor:
\begin{align}
    \nabla_a u_b &= - \mathcal{A} u_a e_b + \left( \frac{\Theta}{3} + \Sigma \right) e_a e_b + \left( \frac{\Theta}{3} - \frac{\Sigma}{2} \right) N_{ab}
    \nonumber \\ & \quad
    + \Omega \eta_{ab}
    \ , \\
    \nabla_a e_b &= - \mathcal{A} u_a u_b + \left( \frac{\Theta}{3} + \Sigma \right) e_a u_b + \frac{\phi}{2} N_{ab} + \xi \eta_{ab}
    \ , \\
    C_{abcd} &= -2 u_a E_{b[c} u_{d]} + 2 u_b E_{a[c} u_{d]} - 2 \eta_{ab}{}^{e} H_{e[c} u_{d]} 
    \nonumber \\ & \quad
    - 2 \eta_{cd}{}^{e} \overline{H}_{e[a} u_{b]} - \eta_{abp} \eta_{cdq} E^{qp}
    \ , \\
    E_{ab} &= u^c u^d C_{acbd} = \mathcal{E} \left( e_a e_b - \frac{N_{ab}}{2} \right) + \tilde{\mathcal{E}} \eta_{ab} 
    \ , \\
    H_{ab} &= \frac{1}{2}\eta_a{}^{pq} C_{pqbr} u^r = \mathcal{H}_r e_a e_b + \frac{1}{2} \mathcal{H}_t N_{ab}
    \ , \\
    \overline{H}_{ab} &= \frac{1}{2} \eta_a{}^{rs} C_{bqrs} u^q = \overline{\mathcal{H}}_r e_a e_b + \frac{1}{2} \overline{\mathcal{H}}_t N_{ab} 
    \ .
\end{align}
We call $\{ \mathcal{A}, \Theta, \Sigma, \Omega, \phi, \xi \}$  {\it kinematic variables} and $\{ \mathcal{E}, \tilde{\mathcal{E}}, \mathcal{H}_r, \mathcal{H}_t, \overline{\mathcal{H}}_r, \overline{\mathcal{H}}_t \}$  {\it Weyl variables}. The $1+1+2$ formalism for non-LRS spacetimes with torsion is fully described in Ref. \citep{CarloniLuz2019}.

The canonical energy-momentum tensor (under local rotational symmetry) is decomposed as
\begin{align}
    S_{ab} 
    &= \mu u_a u_b + p (e_a e_b + N_{ab}) + \Pi \left( e_a e_b - \frac{N_{ab}}{2} \right) 
    \nonumber \\ & \quad
    + 2 Q e_{(a} u_{b)} + 2\tilde{Q} e_{[a} u_{b]} + M \eta_{ab}
    \ .
    \label{CanonicalEnergyMomentuTmensor-LRS-symmetry}
\end{align}

For a Weyssenhoff fluid, we can simplify the decomposition of $S_{ab}$, $\Delta^a{}_{bc}$, and  $T^a{}_{bc}$ collectively. For this, we choose the timelike congruence to be the fluid 4-velocity described in Section \ref{ECWFTheory}. The trade-off is that making this choice fixes the frame, an aspect which will be important in the following discussion. Now, the spin density tensor $L_{ab}$ is decomposed as
\begin{align}
    L_{ab} = \tau \eta_{ab} \ . \label{SpinAngMom-GF-LRS}
\end{align}
Utilising \eqref{SpinAngMom-GF-LRS} and the postulated canonical energy-momentum tensor $S_{ab}$ of the Weyssenhoff fluid \eqref{WeysennFluidEMTnsr}, one obtains
\begin{align}
    S_{ab} &= \mu u_a u_b + p (e_a e_b + N_{ab})
    \ . \label{EnergyMometum-RF-LRS}
\end{align}
From \eqref{SolofHyperMomeq}, the torsion tensor is
\begin{align}
    T^a{}_{bc} = 2\Delta^a{}_{bc} = 2\tau u^a \eta_{bc} \ . \label{SimplifiedTorsionofTheory}
\end{align}
Thus,  $\tau$ simultaneously determines the hypermomentum tensor (a property of matter) and the torsion tensor (a property of the manifold). Collectively, we call $\{ \mu,p,\tau \}$ {\it matter variables}. The kinematic, Weyl, and matter variables together form the full collection of \textit{covariant variables}, which determine the properties of the TLRS spacetime.

Using \eqref{metriccanonicalEMrelation}, we can characterize covariantly the relation between the canonical energy-momentum tensor $S_{ab}$ and the metric energy-momentum tensor $s_{ab}$ as
\begin{align}
    s^{ab} &=  S^{ab} -4\Omega\tau u^a u^b - 2 \Omega\tau N^{ab} + 4\xi\tau u^{(a}e^{b)} \ . \label{metricEMtnsr}
\end{align}
Following \eqref{CanonicalEnergyMomentuTmensor-LRS-symmetry}, $s_{ab}$ is decomposed as
\begin{align}
    s_{ab} &= \overline{\mu} u_a u_b + \overline{p} \left( e_a e_b + N_{ab} \right) + 2 \overline{q} u_{(a} e_{b)}
    \nonumber \\ &  \quad
    + \overline{\Pi} \left( e_a e_b - \frac{N_{ab}}{2} \right)  \ , \label{projection-of-metricEM}
\end{align}
where ${\{ \overline{\mu}, \overline{p}, \overline{\Pi}, \overline{q}\}}$ are the covariant variables defined via projections of the metric energy-momentum tensor. Using equations \eqref{EnergyMometum-RF-LRS}, \eqref{metricEMtnsr} and \eqref{projection-of-metricEM}, we obtain the relations
\begin{equation}
\begin{aligned}
    \overline{\mu} &= \mu - 4\Omega\tau
    \ , &
    \overline{p} &= p - \frac{4}{3}\Omega\tau
    \ , \\
    \overline{\Pi} &= \frac{4}{3}\Omega\tau
    \ , &
    \overline{q} &= 2\xi\tau 
    \ ,
\end{aligned}
\label{metric-canonical-LRSrelations}
\end{equation}
so that the metric energy-momentum tensor for Weyssenhoff fluid with torsion is, in general, not perfect. This aspect will be important in the following discussion.

The absence of matter, i.e. ${L_m = 0}$, leads to 
\begin{equation}\label{True_vacuum}
    {s_{ab} = 0 = \Delta^m{}_{ab}}\ ,
\end{equation} 
since these quantities are directly connected to the matter Lagrangian $L_m$.  In this case, equation \eqref{MetCanonEMRelBad} shows that the canonical energy-momentum tensor vanishes:
\begin{equation}
  S_{ab} = 0\ .
\end{equation}
However, the converse is not true: a vanishing canonical energy-momentum tensor does not necessarily imply \eqref{True_vacuum}. In fact, assuming
\begin{align}
    S_{ab} = 0 \ \implies \ \mu = p=0 \ , \label{Canonical_vacuum}
\end{align}
we have that 
\begin{equation}\label{ProjMetricEMCanVacuum}
\begin{aligned}
    \overline{\mu} &= - 4\Omega\tau
    \ , &
    \overline{p} &= - \frac{4}{3}\Omega\tau
    \ , \\
    \overline{\Pi} &= \frac{4}{3}\Omega\tau
    \ , &
    \overline{q} &= 2\xi\tau 
    \ .
    \end{aligned}
\end{equation}
We call condition \eqref{Canonical_vacuum}  \textit{canonical vacuum}. It is evident that this case does not represent the absence of matter fluid. In a canonical vacuum, the Ricci tensor vanishes, and the only contribution to the Riemann curvature tensor comes from the Weyl tensor. In particular, the investigation of singularity theorems for such a system could lead to interesting results given the important role of energy conditions in the singularity theorems \citep{Curiel_2017}.

\section{Covariant 1+1+2 Equations For TLRS Spacetimes} \label{TLRSEqnDerivation}

The governing equations for the covariant variables are derived via the projections of equations \eqref{BianchiIdType1}, \eqref{oncecontractedBianchiId}, \eqref{ConservationEqn}, and the following Ricci identities of congruences $u^a$ and $e^a$
\begin{align}
    R_{abcd}u^d &= \nabla_a \nabla_b u_c - \nabla_b \nabla_a u_c + T^{k}{}_{ab} \nabla_k u_c
    \label{RicciIdentityU}
     \ , \\
    R_{abcd}e^d &= \nabla_a \nabla_b e_c - \nabla_b \nabla_a e_c + T^{k}{}_{ab} \nabla_k e_c
    \label{RicciIdentityE}
     \ .
\end{align}
In the following Sections \ref{TLRSEqnDerivationRicci}-\ref{TLRSEqnDerivationBianchiII}, we provide a pedagogical derivation of the equations governing the TLRS spacetimes. Then, in Section \ref{FoliationoManifoldConsistencyConditions}, we show how these equations can be utilized to derive some additional constraints. We also present the conditions under which a manifold admits a foliation. The insights from the results of Section \ref{FoliationoManifoldConsistencyConditions} help us to develop a better understanding of the governing equations derived in Sections \ref{TLRSEqnDerivationRicci}-\ref{TLRSEqnDerivationBianchiII}. Subsequently, after performing certain simplifications, the complete set of equations is presented in Section \ref{FullTLRSEqn}.

\subsection{Ricci Identities} \label{TLRSEqnDerivationRicci}
Due to the symmetries of the Riemann tensor, there are only six non-vanishing projections of Ricci identities. Further, two of the projections of Ricci identity of $e^a$ \eqref{RicciIdentityE} are not independent from projections of Ricci identity of $u^a$ \eqref{RicciIdentityU}. The projections of \eqref{RicciIdentityU} along 
\begin{align*}
    \{ u^a h^{bc}, u^a N^{bc}, u^a \eta^{bc}, e^a N^{bc}, \eta^{abc}, \eta^{ab} e^c \}
\end{align*}
are, respectively,
\begin{equation}
\begin{aligned}
    \dot{\Theta} - \hat{\mathcal{A}} &= \mathcal{A} \left( \mathcal{A}+\phi \right) - \frac{\mu}{2} - \frac{3p}{2} + 2\Omega^2 - \frac{3}{2} \Sigma^2 - \frac{1}{3} \Theta^2
    \ , \\
    \dot{\Sigma} - \frac{2}{3} \dot{\Theta} &= - \mathcal{A}\phi - \mathcal{E} + \frac{\mu}{3} + p + \frac{1}{2}\left( \Sigma - \frac{2}{3}\Theta \right)^2 - 2\Omega^2
    \ , \\
    \dot{\Omega} &= \left( \Sigma - \frac{2}{3} \Theta \right) \Omega + \xi \mathcal{A} - \tilde{\mathcal{E}}    
    \ , \\
    \hat{\Sigma} - \frac{2}{3} \hat{\Theta} &= -\frac{3}{2} \phi\Sigma - 2\xi\Omega
    \ , \\
    \hat{\Omega} &= \left( \mathcal{A}-\phi \right) \Omega - \mathcal{A}\tau + \frac{\mathcal{H}_r}{2} + \frac{\mathcal{H}_t}{2}
    \ , \\
    \mathcal{H}_r &= 3\xi\Sigma - \left( 2\mathcal{A} - \phi \right) \Omega  + 2\mathcal{A}\tau
    \ .
\end{aligned}
\label{ProjectedRicciIdentityofUFinish}
\end{equation}
The projections of \eqref{RicciIdentityE} along
\begin{align*}
    \{ u^a N^{bc}, e^a N^{bc}, u^a \eta^{bc}, e^a \eta^{bc} \}
\end{align*}
are, respectively,
\begin{equation}
\begin{aligned}
    \dot{\phi} &= -\left( \frac{\Sigma}{2} - \frac{\Theta}{3} \right) \left( 2\mathcal{A}-\phi \right) + 2\xi\Omega
    \ , \\
    \hat{\phi} &= -\frac{1}{2} \phi^2 + 2\xi^2 + \frac{2}{9} \Theta^2 + \frac{1}{3} \Sigma\Theta - \Sigma^2 - \mathcal{E} - \frac{2}{3} \mu
    \ , \\
    \dot{\xi} &= \left( \mathcal{A} - \frac{\phi}{2} \right)\Omega + \xi\left( \frac{\Sigma}{2} - \frac{\Theta}{3} \right) - \frac{\overline{\mathcal{H}}_t}{2}
    \ , \\
    \hat{\xi} &= \Omega \left( \Sigma + \frac{\Theta}{3} \right) - \xi\phi - \tilde{\mathcal{E}}
    \ .
\end{aligned}
\end{equation}

\subsection{Bianchi Identity Type-I} \label{TLRSEqnDerivationBianchiI}
There are eight independent projections of Bianchi identity type-I \eqref{BianchiIdType1}. However, since \eqref{SimplifiedTorsionofTheory} holds for the Weyssenhoff fluid, we are left with only five independent equations. 
The projections of \eqref{BianchiIdType1} along 
\begin{align*}
\{ \eta^{ab}\delta^c_n, \eta^{ab} u_n e^c, \eta^{ab} e_n e^c, \eta^{ab} e_n u^c, u^a e^b \eta^c{}_n \}
\end{align*}
give the following five equations, respectively,
\begin{equation}
\begin{aligned}
    \dot{\tau} &= -\tau\Theta
    \ , \\    
    \hat{\tau} &= -\tau\phi + \frac{\mathcal{H}_r}{2} + \frac{\mathcal{H}_t}{2}
    \ , \\
    \tilde{\mathcal{E}} &= \tau \left( \Sigma + \frac{\Theta}{3} \right)
    \ , \\
    \overline{\mathcal{H}}_t &= 2 \mathcal{A} \tau - \mathcal{H}_r
    \ , \\
    2 \overline{\mathcal{H}}_r &= - \left(\overline{\mathcal{H}}_t + {\mathcal{H}}_t\right)
    \ .
\end{aligned}
\label{BianchiINewEqns}
\end{equation}
The projections of \eqref{BianchiIdType1} along 
$$\{ u^a e^b \delta^c_n, \eta^{ab} u_n u^c, \eta^{ab} \eta^c{}_n \}$$ 
 do not give any independent equation due to the form of torsion. Equations \eqref{BianchiINewEqns} are crucial in order to understand how the conformal structure of the manifold relates to the kinematic variables and torsion. 

As a side note, we remark that in General Relativity (without torsion), equations \eqref{BianchiINewEqns} do not appear explicitly as they just imply that there is a unique magnetic part of the Weyl tensor (${H_{ab}=\overline{H}_{ab}}$) and that $H_{ab}$ and $E_{ab}$ are traceless and symmetric.

\subsection{Bianchi Identity Type-II} \label{TLRSEqnDerivationBianchiII}
There are clearly two independent projections of \eqref{ConservationEqn}, one along $u^a$ and one along $e^a$ which, respectively, give
\begin{equation}
\begin{aligned}
    \dot{\mu} &= -\Theta \left( \mu + p \right)
    \ , \\
    \hat{p} &= -\mathcal{A} \left( \mu + p \right) -2\tau \overline{\mathcal{H}}_r
    \ .
\end{aligned}
\end{equation}

Finally, the six independent projections of \eqref{oncecontractedBianchiId} are
\begin{align*}
\{ \eta^{ab} e^c, \eta^{ab} u^c, u^a e^b e^c, u^a e^b u^c, u^a \eta^{bc}, e^a\eta^{bc} \}    
\end{align*}
and projections along them give, respectively,
\begin{equation}
\begin{aligned}
\dot{\mathcal{H}}_r 
    &= \left( \frac{\Sigma}{2} - \frac{\Theta}{3} \right) \left( 2\mathcal{H}_r - \overline{\mathcal{H}}_t \right) - \tilde{\mathcal{E}}\phi - 3\xi\mathcal{E}
\ , \\
\hat{\mathcal{H}}_r 
    &= \frac{\phi}{2} \left( \mathcal{H}_t - 2\mathcal{H}_r \right) - \left( \Sigma - \frac{2}{3}\Theta \right)\tilde{\mathcal{E}}  
    \\ & \quad
    - \left( 3\Omega - 2\tau \right) \mathcal{E} + \frac{1}{3}\mu\tau - \mu\Omega - p\Omega + p\tau
\ , \\
\dot{\mathcal{E}} - \frac{\dot{\mu}}{3} 
    &= \left( \frac{3}{2}\Sigma - \Theta \right)\mathcal{E} - \left( \frac{\Sigma}{2} - \frac{\Theta}{3} \right) \left( \mu + p \right) 
    \\ & \quad
    + 2\Omega\tilde{\mathcal{E}} + \xi \left( 2\overline{\mathcal{H}}_r - \overline{\mathcal{H}}_t \right)
\ , \\
\hat{\mathcal{E}} - \frac{\hat{\mu}}{3} 
    &= -\frac{3}{2} \mathcal{E} \phi + \left( 2\overline{\mathcal{H}}_r - \mathcal{H}_t \right) \Omega - 2 \xi \tilde{\mathcal{E}} - 2 \tau \overline{\mathcal{H}}_r
\ , \\
2\dot{\tilde{\mathcal{E}}} - \hat{\overline{\mathcal{H}}}_t 
    &= \mathcal{A} \left( \overline{\mathcal{H}}_t - \mathcal{H}_t \right) - \frac{\phi}{2} \left( 2\overline{\mathcal{H}}_r - \overline{\mathcal{H}}_t \right)  
    \\ & \quad
    -3\mathcal{E}\Omega - \tilde{\mathcal{E}} \left( 3\Sigma + 2\Theta \right) - \Omega \left( \mu+p \right)
\ , \\
2\hat{\tilde{\mathcal{E}}} + \dot{\mathcal{H}}_t 
    &= 3\xi\mathcal{E} - \left( 4\mathcal{A} + \phi \right) \tilde{\mathcal{E}} - \Theta\mathcal{H}_t + \frac{3}{2} \Sigma \overline{\mathcal{H}}_t 
    \\ & \quad
    + \left( \frac{\Sigma}{2} - \frac{\Theta}{3} \right) \left( 4\mathcal{H}_r + 3\mathcal{H}_t + 3\overline{\mathcal{H}}_t + 2\overline{\mathcal{H}}_r \right)
\ .
\end{aligned}
\end{equation}

\subsection{Foliation of Manifold and Consistency Conditions} \label{FoliationoManifoldConsistencyConditions}
To better understand the covariant equations for TLRS spacetimes, we first describe the condition for the foliation of the manifold. Here, we only want to study if the manifold can be foliated into a one-form $n_a$ and a ${3-}$dimensional integral submanifold. 

Let $T\mathcal{M}$ be the tangent space and $T^*\mathcal{M}$ be the co-tangent space on the manifold $\mathcal{M}$. Let $n_a$ be a one-form (which spans a ${1-}$dimensional co-tangent subspace) and ${V \subset T\mathcal{M}}$ be a ${3-}$dimensional tangent subspace complementary to it, that is
\begin{align}
    n_a X^a = 0 \ , \forall X \in V \ ,
\end{align}
then the condition such that tangent subspace $V$ admits an integral submanifold is given by the Frobenius' Theorem (dual formulation) \citep{WaldBook,NomizuInterscience,LuzHypersurface} as follows
\begin{align}
    2n_{[a} \nabla_b n_{c]} + n_{[a} T^{k}{}_{bc]} n_k = 0
    \ .
    \label{OrthogonalityCondition}
\end{align}
This condition originates by demanding that the tangent subspace $V$ must be \textit{involutive}, or equivalently, by demanding that the co-tangent subspace spanned by the one-form $n_a$ must be \textit{differential} \citep{WaldBook,NomizuInterscience,LuzHypersurface}. A tangent subspace $V \subset T\mathcal{M}$ is considered to be involutive if the Lie transport of any vector field $Y \in V$ with respect to vector field $X \in V$ leaves the resulting vector field on the tangent subspace $V$. A co-tangent subspace is said to be differential if it is spanned by a complete set of (generating) one-forms that are parallel to the derivative of a coordinate function. For the $1-$dimensional co-tangent subspace spanned by $n_a$ as given above, being differential implies, in abstract notation, that
\begin{align}
    \textbf{n} = \psi d\varphi \ ,
\end{align}
where \textbf{n} is the one-form with components $n_a$, $\psi$ is a generic function on the manifold, and the function $\varphi$ describes a coordinate function. These conditions are further described in Appendix \ref{Appen-IntegrableSubmanifold}.

In this paper, we refer to condition \eqref{OrthogonalityCondition} as \textit{Hypersurface Orthogonality Condition}.
Further, if a one-form $n_a$ satisfies the hypersurface orthogonality condition \eqref{OrthogonalityCondition}, then the one-form $n_a$ (and its dual $n^a$) is said to be \textit{hypersurface orthogonal} (to the family of integral submanifolds formed by the complementary tangent subspace $V$), and the manifold is said to admit a foliation orthogonal to $n_a$.

Using the hypersurface orthogonality condition and the torsion \eqref{SimplifiedTorsionofTheory}, we see that $u_a$ is hypersurface orthogonal if
\begin{align}
    \Omega - \tau = 0 \ ,
\end{align}
and $e_a$ is hypersurface orthogonal if
\begin{align}
    \xi = 0 \ .
\end{align}
Taking the $\eta^{ab}$ projection of the definition of torsion \eqref{DefofTorsion}, we have
\begin{align}
4\xi \hat{\psi} - 4\Omega\dot{\psi} = -\eta^{ab} T^{k}{}_{ab} \nabla_k \psi \ ,
\end{align}
and further using \eqref{SimplifiedTorsionofTheory}, we obtain
\begin{align}
    (\Omega - \tau) \dot{\psi} = \xi \hat{\psi} \ . \label{DerCons}
\end{align}
Eq.\eqref{DerCons} is useful as it can be imposed on variables which have both an evolution equation ({\it dot} derivative) and a propagation equation ({\it hat} derivative). This process leads to some additional constraints, which are typically referred to as \textit{consistency conditions}. Additionally, \eqref{DerCons} highlights some other properties resulting from the foliation of the manifold. For example, for a spacetime manifold which only admits a foliation orthogonal to a timelike vector field $u^a$, while the $1-$dimensional cotangent subspace spanned by the spacelike vector field $e_a$ is not differential, then this implies that {\it hat} derivative of all scalars must vanish.

Now imposing \eqref{DerCons} for $\xi$ and ${(\Omega-\tau)}$, and using equations from Section \ref{TLRSEqnDerivationRicci} and Section \ref{TLRSEqnDerivationBianchiI}, we get the consistency conditions
\begin{align}
    \xi \beta = 0 = (\Omega-\tau) \beta \ , \label{betaintroduced}
\end{align}
where, following \cite{EllisElst}, we defined
\begin{align}
    \beta \coloneqq \xi\phi + \left( \Omega-\tau \right) \left( \Sigma - \frac{2}{3} \Theta \right) \ .
\end{align}
The solution of \eqref{betaintroduced} either requires ${\beta=0}$ or otherwise, we need ${\xi = 0 =\Omega-\tau}$ which then still implies ${\beta=0}$. Thus, we can replace \eqref{betaintroduced} with simply the solution
\begin{align}
    \beta = 0 \ . \label{betaequation}
\end{align}
The remaining consistency conditions can be derived by utilising the {\it dot} and {\it hat} derivative equations in Sections \ref{TLRSEqnDerivationRicci}, \ref{TLRSEqnDerivationBianchiI}, \ref{TLRSEqnDerivationBianchiII} for the variables
\begin{align*}
    \{ \tau, {(\mathcal{E}-\mu/3)}, \mathcal{H}_r, {(\Sigma - 2\Theta/3)}, \phi \} \ .
\end{align*}
These consistency conditions are \eqref{OtherConsistencyEqn1}, \eqref{AddCons-E}, \eqref{AddCons-H}, \eqref{OtherConsistencyEqn2}, \eqref{OtherConsistencyEqn3} respectively and are reported in the next Section \ref{FullTLRSEqn}.

\subsection{The Complete Covariant Equations for TLRS Spacetimes with Weyssenhoff Fluid} \label{FullTLRSEqn}
The covariant variables which completely describe the TLRS spacetimes with Weyssenhoff fluid (as the matter source) are 
\begin{align}
    \{ \mathcal{A}, \Theta, \Sigma, \Omega, \phi, \xi, \mu, p, \tau, \mathcal{E}, \tilde{\mathcal{E}}, \mathcal{H}_r, \mathcal{H}_t, \overline{\mathcal{H}}_r, \overline{\mathcal{H}}_t \} \ .
\end{align}
We utilize \eqref{betaequation}, \eqref{BianchiINewEqns} and the last equation of \eqref{ProjectedRicciIdentityofUFinish} to simplify the equations derived in Sections \ref{TLRSEqnDerivationRicci}, \ref{TLRSEqnDerivationBianchiI}, \ref{TLRSEqnDerivationBianchiII}, \ref{FoliationoManifoldConsistencyConditions} and to replace the propagation and evolution equations of $\Omega$ with the ones for ${(\Omega-\tau)}$. In this way, we collect and present the final version of all the covariant equations for the TLRS spacetimes for Einstein-Cartan theory with Weyssenhoff fluid. 

\textit{Evolution Equations:}
\begin{align}
    \dot{\Omega} - \dot{\tau} &= \xi \left( \mathcal{A}-\phi \right)
    \label{omegataudot} \\
    \dot{\xi} &= \xi \left( 2\Sigma - \frac{\Theta}{3} \right)
    \label{xidot} \\
    \dot{\tau} &= -\tau\Theta
    \label{TauDot}
    \\
    \dot{\Sigma} - \frac{2}{3} \dot{\Theta} &= 
        - \mathcal{A}\phi - \mathcal{E} + \frac{\mu}{3} + p + 2\left( \frac{\Sigma}{2} - \frac{\Theta}{3} \right)^2 - 2\Omega^2
    \label{chidot}
    \\
    \dot{\phi} &= - \left( \frac{\Sigma}{2} - \frac{\Theta}{3} \right) \left( 2\mathcal{A}-\phi \right) + 2\xi\Omega \label{DotPhi} 
    \\
    \dot{\mathcal{E}} - \frac{\dot{\mu}}{3} &= 
            \left( \frac{\Sigma}{2} - \frac{\Theta}{3} \right) (3\mathcal{E} - \mu - p)
            + 2\Omega\tilde{\mathcal{E}} 
            \nonumber \\ & \quad
            + \xi \left( 2\overline{\mathcal{H}}_r - \overline{\mathcal{H}}_t \right)
    \label{ElecEnergydot} \\
    \dot{\mathcal{H}}_r &= 
        \left( \frac{\Sigma}{2} - \frac{\Theta}{3} \right) \left( 2\mathcal{H}_r - \overline{\mathcal{H}}_t \right) 
        - \tilde{\mathcal{E}}\phi - 3\xi\mathcal{E} 
    \label{Redun1}
    \\
    \dot{\mu} &= -\Theta \left( \mu + p \right)
    \label{energydot}
\end{align}

\textit{Propagation Equations:}
\begin{align}
    \hat{\Omega} - \hat{\tau} &= \left( \Omega-\tau \right) \left( \mathcal{A}-\phi \right)
    \label{omegatauhat} \\
    \hat{\xi} &= \left( \Omega-\tau \right) \left( 2\Sigma - \frac{\Theta}{3} \right)
    \label{xihat} \\
    \hat{\tau} &= -\tau\phi + \frac{\mathcal{H}_r}{2} + \frac{\mathcal{H}_t}{2}
    \label{TauHat}
    \\
    \hat{\Sigma} - \frac{2}{3} \hat{\Theta}
        &= -\frac{3}{2} \phi\Sigma - 2\xi\Omega
    \label{chihat} \\
    \hat{\phi} &= -\frac{1}{2} \phi^2 + 2\xi^2 + \frac{2}{9}\Theta^2 + \frac{1}{3}\Sigma\Theta - \Sigma^2 - \mathcal{E} - \frac{2}{3}\mu
    \label{phihat} \\
    \hat{\mathcal{E}} - \frac{\hat{\mu}}{3} &= 
            -\frac{3}{2} \mathcal{E} \phi 
            + \left( 2\overline{\mathcal{H}}_r - \mathcal{H}_t \right) \Omega 
            - 2 \xi \tilde{\mathcal{E}} 
            - 2 \tau \overline{\mathcal{H}}_r
    \label{ElecEnergyhat} \\
    \hat{\mathcal{H}}_r &= -\frac{\phi}{2} \left( 2\mathcal{H}_r - \mathcal{H}_t \right) 
                            - \left( \Sigma - \frac{2}{3} \Theta \right) \tilde{\mathcal{E}} 
                            \nonumber \\ & \quad
                            - \left( 3\Omega - 2\tau \right) \mathcal{E}
                            +\frac{1}{3}\mu\tau - \mu\Omega - p\Omega + p\tau
    \label{Redun2}
    \\
    \hat{p} &= -\mathcal{A} \left( \mu + p \right) -2\tau \overline{\mathcal{H}}_r
    \label{preshat}
\end{align}

\textit{Mixed Equations:}
\begin{align}
    \dot{\Theta} - \hat{\mathcal{A}} &= \mathcal{A} \left( \mathcal{A}+\phi \right) - \frac{\mu}{2} - \frac{3}{2}p + 2\Omega^2 - \frac{3}{2} \Sigma^2 - \frac{1}{3}\Theta^2
    \\
    2\hat{\tilde{\mathcal{E}}} + \dot{\mathcal{H}}_t &= 
            3\xi\mathcal{E} - \left( 4\mathcal{A} + \phi \right) \tilde{\mathcal{E}} - \Theta\mathcal{H}_t 
            + \frac{3}{2} \Sigma \overline{\mathcal{H}}_t
            \nonumber \\ & \quad
            + \left( \frac{\Sigma}{2} - \frac{\Theta}{3} \right) \left( 4\mathcal{H}_r + 3\mathcal{H}_t + 3\overline{\mathcal{H}}_t + 2\overline{\mathcal{H}}_r \right)
    \label{EtildehatHtdoteqn} \\
    2\dot{\tilde{\mathcal{E}}} - \hat{\overline{\mathcal{H}}}_t &= 
            \mathcal{A} \left( \overline{\mathcal{H}}_t - \mathcal{H}_t \right) 
            - \frac{\phi}{2} \left( 2\overline{\mathcal{H}}_r - \overline{\mathcal{H}}_t \right)
            -3\mathcal{E}\Omega 
            \nonumber \\ & \quad
            - \tilde{\mathcal{E}} \left( 3\Sigma + 2\Theta \right) - \Omega \left( \mu+p \right)
    \label{Redun3}
\end{align}

\textit{Constraint Equations:}
\begin{align}
    \tilde{\mathcal{E}} &= \tau \left( \Sigma + \frac{\Theta}{3} \right)
    \label{ETilde-Constraint}
    \\
    \mathcal{H}_r &= 3\xi\Sigma - \left( 2\mathcal{A} - \phi \right) \Omega  + 2\mathcal{A}\tau
    \label{Hr-Constraint}
    \\
    \overline{\mathcal{H}}_t &= 2\mathcal{A}\tau - \mathcal{H}_r = -3\xi\Sigma + \left( 2\mathcal{A} - \phi \right) \Omega
    \label{Htbar-Constraint}
    \\
    2\overline{\mathcal{H}}_r &= - \overline{\mathcal{H}}_t - \mathcal{H}_t
    \label{HrbarHt-Constraint}
\end{align}

\textit{Consistency Conditions:}
\begin{align}
    \beta &= 0 = \xi\phi + \left( \Sigma - \frac{2}{3}\Theta \right) \left( \Omega-\tau \right) 
    \label{thebeta}
    \\
    \xi \mathcal{H}_t &= -\xi\mathcal{H}_r - 2\tilde{\mathcal{E}} \left( \Omega-\tau \right)
    \label{OtherConsistencyEqn1}
    \\
    0 &= 
    \left( \frac{\Sigma}{2} - \frac{\Theta}{3} \right) \left( \Omega-\tau \right) \left( \mu+p \right) 
    \nonumber \\ &\qquad
    - \left( \Sigma - \frac{2}{3}\Theta \right) \tau \left( \Omega^2 - \Omega\tau - \xi^2 \right)
    \label{AddCons-E} 
    \\
    0 &=
    \xi \left( \Omega-\tau \right) \left( \mu+p \right) 
    \nonumber \\ & \qquad
    - 2\xi\tau \left( \Omega^2 - \Omega\tau - \xi^2 \right) 
    \label{AddCons-H} 
    \\
    2 \xi^2 \tau &=
    \left( \Omega-\tau \right) \left\{ \mathcal{E} + \mathcal{A}\phi +  \left( \Sigma - \frac{2}{3}\Theta \right) \left( \Sigma+\frac{\Theta}{3} \right) 
    \right. \nonumber \\ & \qquad \qquad \qquad \left.
    + 2\Omega^2 - 2\xi^2 - \frac{\mu}{3} - p \right\}
    \label{OtherConsistencyEqn2}
    \\
    2 \xi \Omega \tau &=
    \xi \left\{ \mathcal{E}  + \mathcal{A}\phi + \left( \Sigma - \frac{2}{3}\Theta \right) \left( \Sigma+\frac{\Theta}{3} \right) 
    \right. \nonumber \\ & \qquad \qquad \qquad \left.
    + 2\Omega^2 - 2\xi^2 + \frac{2}{3}\mu \right\}&
    \label{OtherConsistencyEqn3}
\end{align}
Equations \eqref{Redun1}, \eqref{Redun2} and \eqref{Redun3} are redundant since they can be obtained by taking the derivative of relevant constraint equations (\ref{ETilde-Constraint}-\ref{HrbarHt-Constraint})\footnote{Equations \eqref{AddCons-E} and \eqref{AddCons-H}, as presented here, have been simplified using equations \eqref{OtherConsistencyEqn1} and \eqref{OtherConsistencyEqn3}.}. The above set of equations is closed when one has enough information on the matter fluid, in general, given by an equation of state ${p=p(\mu)}$ and a relation between $\tau$ and $\mu$ that we will call \textit{equation of spin density}. However, the necessary information can be provided in different forms or might not be required for certain systems (for example, canonical vacuum, see Section \ref{TLRS1B-CanonicalVacuum}). The equation of state can also be provided in terms of projections of the metric energy-momentum tensor as ${f(\overline{\mu},\overline{p},\overline{\Pi})=0}$. Similarly, the information on $\tau$ can be provided via a matter Lagrangian and the resulting hypermomentum tensor derived from its variation with respect to the contortion tensor (see Ref. \citep{Hehl,Capozziello,Korotky}).
Also, notice that the role of consistency conditions depends heavily on the foliation the manifold admits. Indeed, most known exact solutions possess both a timelike and a spacelike hypersurface orthogonal one-forms, and for such spacetimes, the consistency conditions vanish identically.

\section{Classification of LRS Spacetime in General Relativity}
\label{Classification-of-LRSGR-Subsection}
In this section, we review the classification of LRS spacetimes in General Relativity \citep{EllisElst} and highlight some salient features of the classification. These results will be useful to form the basis for the classification of TLRS spacetimes described in Section \ref{Classification-of-TLRSSpacetimes-subsection}.

First of all, we notice that the 1+1+2 equations given in the previous section reduce to the GR ones by setting ${\tau=0}$. Setting also $\Pi=Q=0$, so that matter is modelled as a perfect fluid, i.e., there is no energy flux or anisotropic pressure\footnote {Notice here that equation \eqref{EnergyMometum-RF-LRS} already represents the canonical energy-momentum tensor of a perfect fluid. So, in the case of ECSK with Weyssenhoff fluid, this step is not really required.}, equation \eqref{AddCons-H} becomes \citep{EllisElst}
\begin{align}\label{PreOmegaxiConsitutiveEqn}
    \xi \Omega \left( \mu + p \right) &= 0 \ .
\end{align}
It is important to stress that one needs to impose the condition ${\mu+p \neq 0}$ for the classification to make sense.  This condition is essentially a constraint on the equation of state $p=p(\mu)$, which is always a characteristic property of the fluid, not determined by the Einstein equation. However, in Ref. \citep{EllisElst}, one considers a stricter condition ${\mu+p > 0}$ on the matter variables, which can be understood in various ways.
Thermodynamically, it can be seen as a non-vacuum case with the isentropic speed of sound bounded by the requirement of local stability of matter and causality \citep{EllisElst1998CM}. Furthermore, such a condition can be seen as a strict form of the null energy condition, which would imply the purely attractive nature of gravitational fields \citep{Curiel_2017}. Assuming ${\mu+p \neq 0}$ in \eqref{PreOmegaxiConsitutiveEqn}  leads to the crucial relation \citep{EllisElst}
\begin{align}
    \Omega \xi &= 0 \ . \label{OmegaxiConsitutiveEqn}
\end{align}
which forms the basis of the classification. Under the same conditions, we can also derive the relations \citep{EllisElst}
\begin{align}
\Omega\Theta=\Omega\Sigma=\xi\mathcal{A}=\xi\phi=0 \ . \label{ConstitutiveAll}
\end{align}
The classification of LRS spacetimes is based on the three possible solutions of \eqref{OmegaxiConsitutiveEqn}. These solutions also determine which foliations the manifold admits. In particular, we have
\begin{align*}
    \text{LRS}& \text{  Class I:} \ \Omega\neq 0,\xi=0 \\ & \implies \text{only } e_a \text{ is hypersurface orthogonal,}
    \\
    \text{LRS}& \text{  Class II:} \ \Omega=0,\xi=0  \\ & \implies \text{both } (u_a,e_a) \text{ are hypersurface orthogonal,}
    \\
    \text{LRS}& \text{  Class III:} \ \Omega=0,\xi\neq 0 \\ & \implies \text{only } u_a \text{ is hypersurface orthogonal.}
\end{align*}
We will now proceed to demonstrate another key feature of the classification of LRS spacetimes. To achieve this result, we begin with a local statement on the vanishing of the variables $\Omega$ and $\xi$  and then show that this statement holds true for the entire manifold, i.e., it is global\footnote{Note that since $\Omega$ and $\xi$ are variables defined through the covariant derivative of congruences, local measurements cannot provide information on their local values.}. To achieve this result, we employ the covariant equations of $\Omega$ and $\xi$ for LRS spacetimes, which can be recast as\footnote{These equations can be obtained using equations \eqref{omegataudot}, \eqref{xidot}, \eqref{omegatauhat} \eqref{xihat} with $\tau=0$, or after an equivalent manipulation of the equations given in Ref. \citep{EllisElst}.} 
\begin{equation}
\begin{aligned}
    \dot{\Omega} &= \xi \left( \mathcal{A}-\phi \right)
    \ , &
    \hat{\Omega} &= \Omega \left( \mathcal{A}-\phi \right)
    \ , \\
    \dot{\xi} &= \xi \left( 2\Sigma - \frac{\Theta}{3} \right)
    \ , &
    \hat{\xi} &= \Omega \left( 2\Sigma - \frac{\Theta}{3} \right)
    \ .
\end{aligned}
\label{LRS-Equations}
\end{equation}
Let us start with the simple case of LRS class II spacetimes. Suppose the local value of $\Omega$ and $\xi$ at some point $x\in\mathcal{M}$ to be given as ${\Omega|_x=0}, \ {\xi|_x=0}$. Then from equation \eqref{LRS-Equations}, we can conclude that
\begin{align}
    \dot{\Omega}|_x &= 0 \ , &
    \hat{\Omega}|_x &= 0 \ , &
    \dot{\xi}|_x &= 0 \ , &
    \hat{\xi}|_x &= 0 \ .
\end{align}
This implies that $\Omega=0=\xi$ is true in a neighborhood  $\mathcal{N}_x$ of  $x$. Now, at an arbitrary point $y\in \mathcal{N}_x \ (y\neq x)$, the relations above remain true, and therefore 
\begin{align}
    \dot{\Omega}|_y &= 0 \ , &
    \hat{\Omega}|_y &= 0 \ , &
    \dot{\xi}|_y &= 0 \ , &
    \hat{\xi}|_y &= 0 \ ,
\end{align}
so that  $\Omega$ and $\xi$ vanish in a neighborhood $\mathcal{N}_y$ of the point $y$. As $\mathcal{N}_y$ can extend beyond $\mathcal{N}_x$, by repeating this procedure for other points, we can eventually cover the entire manifold $\mathcal{M}$ with overlapping open sets in which $\Omega=0=\xi$. In this way, $\Omega=0=\xi$ can be recognized as a global property of the spacetime manifold.

The above result implies that if at a point in the spacetime $\Omega=0=\xi$, then this property will be true throughout the manifold, and we will have an LRS class II spacetime. In other words, if $\Omega=0=\xi$ at a point in spacetime, there cannot be any points at which ${\Omega\neq0}$ or ${\xi\neq0}$.  This characteristic feature is codified in the concept of \textit{separation of classes}. We say that an LRS class is \textit{separated} if the characterizing quantities of this class assume globally the same value that they present locally. Using similar arguments, it is not difficult to prove that all the LRS classes are separated.

Let us consider, for example, LRS class III spacetimes. Let the local value of $\Omega$ and $\xi$ at some point $x\in\mathcal{M}$ be given as ${\Omega|_x=0}, \ {\xi|_x\neq0}$. Then from equation \eqref{ConstitutiveAll}, we can conclude that
\begin{align}
\mathcal{A}|_x = 0 = \phi|_x
\end{align}
Substituting in \eqref{LRS-Equations}, we obtain
\begin{equation}
 \dot{\Omega}|_x=\hat{\Omega}|_x = \hat{\xi}|_x =0 \ ,
 \end{equation}
 which ensures that ${\Omega=0}$ everywhere in $\mathcal{N}_x$, and 
\begin{equation}
\dot{\xi}|_x = \xi|_x \left( 2\Sigma - \frac{\Theta}{3} \right)\bigg|_x \label{LRSIII_dot_xi}\ .
 \end{equation}    
Since in a neighborhood  $\mathcal{N}_x$ of $x$, we can consider $\Theta|_x$ and $\Sigma|_x$ to be constant and finite, we can see that \eqref{LRSIII_dot_xi} has an exponential solution. As a consequence, $\xi$ cannot be zero in $\mathcal{N}_x$. Now, taking an arbitrary point $y\in \mathcal{N}_x \ (y\neq x)$, we can show in the same way that ${\Omega=0}, \ {\xi\neq0}$ in a neighborhood $\mathcal{N}_y$ that extends beyond $\mathcal{N}_x$. In this way, as before, we can cover the entire manifold with overlapping open sets, each of which satisfies $\Omega=0$ and $\xi\neq0$, making it a global property of the spacetime manifold. The same reasoning also leads to ${\mathcal{A} = 0 = \phi}$ for entire manifold $\mathcal{M}$ through equation \eqref{ConstitutiveAll}. The entire spacetime, therefore, belongs to LRS class III. 

One can also perform similar steps in the case of spacetimes belonging to LRS class I: starting from ${\Omega|_x\neq0},\ {\xi|_x=0}$ and showing that this property is globally true. We have then proven that in GR, all LRS classes are separated\footnote{Notice that, if one tries to classify the LRS spacetimes based on the solution of any other equation in \eqref{ConstitutiveAll}, then the classification is not possible, without assuming additional global properties on other kinematic variables.}.

While developing the classification for TLRS spacetimes, it will be necessary to ensure that TLRS classes are separated. We will proceed in the same manner as described above: given a local statement on the variables which define each class, we will prove that the same statement holds for the entire manifold.

\section{Classification of TLRS Spacetimes} 
\label{Classification-of-TLRSSpacetimes-subsection}
We first begin with developing a scheme for the classification of TLRS spacetimes (with Weyssenhoff fluid) in the comoving frame based on the results from Section \ref{Classification-of-LRSGR-Subsection}. Then, in the following subsections, we shall describe the TLRS classes and their properties. 

In analogy with the case of General Relativity, one would expect the classification to be based on the quantities $ (\Omega-\tau) $ and $\xi$. This anzats can be supported by two arguments. Firstly, for TLRS spacetimes with Weyssenhoff fluid, the quantities ${(\Omega-\tau)}$ and $\xi$ determine if the spacetime manifold admits a foliation. Secondly, even the covariant equations \eqref{omegataudot}, \eqref{xidot}, \eqref{omegatauhat} and \eqref{xihat} for TLRS spacetimes are similar to the ones for LRS spacetimes \eqref{LRS-Equations} with replacement of ${\Omega \rightarrow (\Omega-\tau)}$.

However, we immediately find a fundamental difference from the case of GR. Indeed, we see that an extension of the relation \eqref{OmegaxiConsitutiveEqn} for torsion spacetimes is absent. This fact can be seen by combining  equations \eqref{omegataudot}, \eqref{xidot}, \eqref{omegatauhat}, \eqref{xihat} and \eqref{thebeta} to obtain:
\begin{align}
    (\Omega - \tau) \dot{\xi} + (\dot{\Omega} - \dot{\tau}) \xi &= \xi \gamma
    \ , \\
    (\Omega - \tau) \hat{\xi} + (\hat{\Omega} - \hat{\tau}) \xi &= (\Omega - \tau) \gamma
    \ ,    
\end{align}
where
\begin{align}
    \gamma = \xi \mathcal{A} + 3 (\Omega - \tau) \left( \Sigma - \frac{\Theta}{3} \right)
    \ .
\end{align}
Since, in general ${\gamma\neq 0}$, we can conclude through proof by negation\footnote{Assuming ${(\Omega - \tau) \xi = 0}$ leads to $\gamma=0$, which is not true in general, thus the assumption must be false.} that for TLRS spacetimes, in general, we have 
\begin{align}
     (\Omega - \tau) \xi \neq 0 \ .
\end{align}

This is not surprising, and the reason why such an extension is absent can be seen from \eqref{AddCons-H}, rewritten as
\begin{align}
\xi (\Omega-\tau)(\mu + p - 2\tau^2) = 2 \xi \tau \left[ (\Omega-\tau)^2 - \xi^2 \right] \ . \label{H-CC-CanEM}
\end{align}
This relation shows that the presence of torsion and the consequent modification to the nature of gravitational fields and the matter fluid (which now carries spin) leads to the absence of a trivial extension of relation \eqref{OmegaxiConsitutiveEqn} to TLRS spacetimes. 

The situation becomes physically clearer if we write the above equation in terms of the projection of the metric energy-momentum tensor \eqref{projection-of-metricEM} 
\begin{align}
    \xi (\Omega-\tau)(\overline{\mu} + \overline{p} +\overline{\Pi} + 2\tau^2) 
    = 
    -\overline{q} \left[ (\Omega-\tau)^2 + \xi^2 \right] \ . \label{H-CC-MetricEM}
\end{align}
The non-vanishing energy flux $\overline{q}$ in the metric energy-momentum tensor induced by the spin of the matter fields breaks the perfect fluid condition and allows for spacetime with $\{\Omega-\tau,\xi\neq0\}$, irrespective of assumptions on the behavior of the matter.

Since the torsion term $\tau$ appears explicitly in the LHS of equations \eqref{H-CC-CanEM} and \eqref{H-CC-MetricEM} in combination with the other matter variables,  any conditions on the matter variables would involve simultaneously constraining the equation of state and the equation of spin density. Such conditions do not necessarily have a strongly motivated physical justification as the (strict) null energy condition ${\mu+p>0}$ required for the GR classification \citep{HehlSingularity, ArminUjjwal}.

In the absence of an extension of \eqref{OmegaxiConsitutiveEqn} to TLRS spacetimes, the classification scheme as presented below is entirely based on the foliation admitted by the manifold. Thus, we will define classes via statements on the variables $\xi$ and ${(\Omega-\tau)}$. 
This scheme is motivated by and draws from the hypersurface orthogonality of congruences in the various classes of LRS spacetimes in General Relativity, which was highlighted in Section \ref{Classification-of-LRSGR-Subsection}.
Another key feature we aim to preserve is the separation of classes. Therefore, we begin with a local value of $\xi$ and ${(\Omega-\tau)}$. We will then proceed to show, by analyzing the covariant equations for TLRS spacetimes given in Section \ref{FullTLRSEqn} and equation \eqref{DerCons}, that those values of $\xi$ and ${(\Omega-\tau)}$ are global and true on the entire manifold.  As in General Relativity, we may need to impose conditions on the equation of state and hypermomentum, i.e., on $\tau$ (see \eqref{SimplifiedTorsionofTheory}). 

\subsection{TLRS Class I}\label{sec-TLRSclassI}
We start with assuming a local value at $x\in\mathcal{M}$ for our classification variables:
\begin{align}
    (\Omega-\tau)|_x \neq 0 \ , &&& \xi|_x=0 \ . \label{C1localvalue}
\end{align}
Substituting \eqref{C1localvalue} in equations \eqref{DerCons}, \eqref{thebeta}  we obtain
\begin{align}
   \dot{\psi}|_x &= 0 \label{C1dotzero} \ , \\
     \Sigma|_x &= \frac{2}{3}\Theta|_x \ , \label{C1SigmaTheta}
\end{align}
where $\psi$ is a generic scalar. Utilising \eqref{C1dotzero} in equations \eqref{TauDot}, \eqref{energydot}, we obtain
\begin{align}
    \tau|_x\Theta|_x &= 0 \ ,
    \label{C1torsionthetazero}\\
    (\mu+p)|_x\Theta|_x &= 0 \ .
    \label{C1matterthetazero}
\end{align}

Now, we wish to see if $\xi$ vanishes globally. We know that $\dot{\xi}|_x$ must vanish due to \eqref{C1dotzero}, and this is confirmed by \eqref{xidot} when we apply \eqref{C1SigmaTheta}. However, \eqref{xihat} gives
\begin{align}
    \hat{\xi}|_x &= \Theta|_x(\Omega-\tau)|_x \ . \label{C1xihatatq}
\end{align}
Thus, if we can ensure that $\Theta|_x$ vanishes, $\hat{\xi}|_x$ would also vanish, and the value of $\xi$ is preserved in a neighborhood $\mathcal{N}_x$ of $x$.  However, since we do not have an equation for $\hat{\Theta}$ \footnote{Equation \eqref{chihat} would also not be useful. Even taking into account  \eqref{C1SigmaTheta} we cannot make any statements on $\hat\Sigma$ which cannot be deduced by any other equation.}, the only way to control the behavior of $\Theta$ is to use \eqref{C1torsionthetazero} or \eqref{C1matterthetazero}. These equations imply the requirement 
\begin{align}
    \qquad \mu+p&\neq0 & &\text{or} & \tau&\neq0 \ . \label{C1matterconditions}
\end{align}
which, like all statements on the properties of matter, are global conditions by definition. It follows that if either of the conditions \eqref{C1matterconditions} is satisfied, then  $\Theta|_x=0$, and this, in turn, preserves the value of $\xi$ in $\mathcal{N}_x$. 

Let us now turn to the behavior of $(\Omega-\tau)$ in $\mathcal{N}_x$. From equation \eqref{omegatauhat}, we have
\begin{align}
    (\hat{\Omega}-\hat{\tau})|_x = (\Omega-\tau)|_x(\mathcal{A}-\phi)|_x \ .
\end{align}
Since we can consider $(\mathcal{A}-\phi)|_x$ to be constant and finite in $\mathcal{N}_x$, in analogy to what we have seen in the context of GR, we know that the above equation has an exponential solution. Thus, since $(\Omega-\tau)|_x\neq0$ from \eqref{C1localvalue}, we can conclude that $(\Omega-\tau)\neq0$ everywhere in $\mathcal{N}_x$.

Now, we can follow the same steps at an arbitrary point $y\in \mathcal{N}_x$ ${(y\neq x)}$. Given that \eqref{C1localvalue} is true at $y$ and remembering that \eqref{C1matterconditions} is a global property of matter, we can conclude that 
\begin{align}
    (\Omega-\tau)\neq 0 \ , &&& \xi=0 \ ,
\end{align}
is true in a generic neighborhood $\mathcal{N}_y$ of $y$. As $\mathcal{N}_y$ can extend beyond $\mathcal{N}_x$, we can obtain a covering of the manifold $\mathcal{M}$ made of open sets, each of which satisfies $(\Omega-\tau)\neq 0$ and $\xi=0$. Hence, \eqref{C1localvalue} is satisfied everywhere in $\mathcal{M}$, making it a global property of the spacetime manifold. 

We can now define TLRS class I as follows:
\begin{definition}
TLRS class I spacetimes are defined, in the comoving frame, by the property 
\begin{align}
\Omega-\tau\neq0 \ , \quad \xi=0  \ .    
\end{align}
The manifold of this class only admits a foliation into a hypersurface orthogonal spacelike vector field and a 3-D Lorentzian submanifold. Further, any spacetime belonging to this class must belong to one of the following subclasses  determined by the matter fluid properties \eqref{C1matterconditions}: 
\begin{enumerate}
    \item {\bf TLRS class IA} The matter fluid must satisfy the condition on the equation of state $p=p(\mu)$ such that 
    \begin{align}
        \mu+p(\mu)\neq0 \ . \label{C1mupnonzero}
    \end{align}
This subclass is an extension of the LRS class I of General Relativity and can be reduced to LRS I if one imposes $\tau=0$. The matter fluid can be of two types: a spin matter fluid with $\mu+p>0$, or a generalization of phantom matter fluid with spin which satisfies $\mu+p<0$ \citep{Dabrowski,LudwickPhantom}. Notice that due to non-vanishing torsion, the condition $\mu+p>0$ does not necessarily have the same physical implications as described in Section \ref{Classification-of-LRSGR-Subsection}.
    \item {\bf TLRS class IB} The matter fluid must satisfy the condition on the hypermomentum 
    \begin{align}
        \tau \neq 0 \ . \label{C1taunonzero}
    \end{align}
    The spacetime of this subclass cannot be reduced to General Relativity. 
\end{enumerate}
\end{definition}
Typically, we expect the condition on the matter fluid to be global. Thus, we shall have either TLRS class IA or class IB spacetimes if the matter fluid satisfies \eqref{C1mupnonzero} or \eqref{C1taunonzero}, respectively. Some spacetimes may satisfy both the matter conditions globally and, therefore, belong to both classes IA and IB.  However, strictly speaking, we only need to ensure that the matter fluid does not violate the conditions $\tau\neq0$, $\mu+p\neq0$ at the same point. We can also have TLRS class I spacetime, which satisfies one condition on a patch and the other one on the complementary patch of global spacetime. Hence, according to our definitions, classes IA and IB are not separated, and for this reason, we place them as {\it subclasses} within TLRS class I. For both subclasses IA and IB, we have $\Theta=0=\Sigma$, and the spacetime is stationary ($\dot{\psi}=0$, for any generic function $\psi$). This also simplifies the Weyl tensor. In particular, we have $\tilde{\mathcal{E}}=0$ for both subclasses.

\subsection{TLRS Class II}\label{sec-TLRSclassII}
Suppose that at $x\in\mathcal{M}$  we have
\begin{align}
    (\Omega-\tau)|_x = 0 = \xi|_x \ . \label{C2localvalue}
\end{align}
Using \eqref{C2localvalue} in equations \eqref{omegataudot}, \eqref{xidot}, \eqref{omegatauhat}, \eqref{xihat}, we obtain
\begin{align}
    (\dot{\Omega}- \dot{\tau})|_x = (\hat{\Omega}- \hat{\tau})|_x = \dot{\xi}|_x = \hat{\xi}|_x = 0 \ , \label{C2localvaluederivative}
\end{align}
so that  $\Omega-\tau=0$ and $\xi=0$ in the neighborhood $\mathcal{N}_x$. Now, we can apply the same argument used in the case of LRS class II spacetimes. We can choose another arbitrary point $y\in \mathcal{N}_x$ ${(y \neq x)}$ and since \eqref{C2localvalue} holds,  we can conclude that both $\xi$ and ${(\Omega-\tau)}$ vanish in $\mathcal{N}_y$. As $\mathcal{N}_y$ can extend beyond $\mathcal{N}_x$, we can form a covering of the manifold composed of a union of overlapping open sets in which $\Omega-\tau=0=\xi$,  making \eqref{C2localvalue} a global property of the manifold.
We can now define TLRS class II as follows:

\begin{definition}
TLRS class II spacetimes are defined, in the comoving frame, by the property 
\begin{align}
\Omega-\tau=0=\xi \ .    
\end{align}
The manifold of this class admits a foliation into a hypersurface orthogonal spacelike vector field, a hypersurface orthogonal timelike vector field, and a 2-D Riemannian submanifold.
\end{definition}

TLRS class II is essentially an extension of the LRS class II of GR and can be reduced to it if one imposes $\tau=0$. Notice that TLRS class II does not require any global condition on the equation of state of the matter fluid. This may seem counter-intuitive, as in Section \ref{Classification-of-LRSGR-Subsection}, we argued that the condition $\mu+p > 0$ is necessary for constructing the LRS classification in GR.  However, in GR, one can discuss a spacetime which admits the same foliation as LRS class II without pre-supposing any condition on the matter variables, thus allowing for richer systems with anisotropic stress in LRS class II (see, e.g., \citep{CarloniTOVAniso1Fluid,CarloniNaidu}). Our classification scheme will naturally enable us to include those spacetimes in the LRSII class. 

Like their GR counterparts, TLRS class II can be used to study complex cosmological models, like Bianchi or Tolman-Bondi models in cosmology, and to develop models of the interior of stellar objects.  There are many examples of such spacetimes in literature, albeit not treated using covariant formalisms. An exception is given by their application in developing models of compact stars composed of spin matter fluid, see \citep{CarloniLuz2019}. We will not further analyze TLRS class II in this paper.

\subsection{TLRS Class III}\label{sec-TLRSclassIII}
We assume that, at $x\in\mathcal{M}$,  
\begin{align}
    (\Omega-\tau)|_x = 0 \ , &&& \xi|_x \neq 0 \ . \label{C3localvalue}
\end{align}
Using \eqref{C3localvalue},  equations \eqref{DerCons}, \eqref{thebeta}  give
\begin{align}
  \hat{\psi}|_x &= 0 \label{C3hatzero}
    \ , \\
    \phi|_x &= 0 \label{C3phizero}  
    \ ,
\end{align}
where $\psi$ is a generic scalar. Substituting  \eqref{C3localvalue}, \eqref{C3hatzero} and \eqref{C3phizero} in equations  \eqref{chihat} and \eqref{preshat},  we obtain
\begin{gather}
    \Omega|_x\xi|_x = 0 \ \xRightarrow[]{} \ \Omega|_x = \tau|_x = 0 \ , \label{C3tauzero}
    \\
    \mathcal{A}|_x(\mu+p)|_x=0 \ . \label{C3matterAzero}
\end{gather}

We now wish to prove that ${(\Omega-\tau)}$ vanishes globally. Equation \eqref{omegatauhat} is trivial at $x$. Looking at equation \eqref{omegataudot}, since \eqref{C3phizero} holds, we have 
\begin{align}
    (\dot{\Omega} - \dot{\tau})|_x = \xi|_x\mathcal{A}|_x \ .
\end{align}
so that $\mathcal{A}|_x=0$ would imply that ${(\Omega-\tau)}$ vanishes in $\mathcal{N}_x$. Since we do not have an equation for  $\dot{\mathcal{A}}$, we cannot control the behavior of the variable ${\mathcal{A}}$. On the other hand, if the matter fluid follows the global condition $\mu+p\neq0$, then equation \eqref{C3matterAzero} allows us to conclude that $\mathcal{A}|_x=0$ using a global statement. Then ${(\Omega-\tau)}$ vanishes in $\mathcal{N}_x$ and from equation \eqref{xidot} we have
\begin{align}
\dot{\xi}|_x = \xi|_x \left(2\Sigma-\frac{\Theta}{3}\right)\bigg|_x \ .
\end{align}
As before, since we can consider $\Theta|_x$ and $\Sigma|_x$ to be constant and finite in $\mathcal{N}_x$, the above equation has an exponential solution, and we can conclude that $\xi$ is non-vanishing in $\mathcal{N}_x$. 

Now take an arbitrary point $y\in \mathcal{N}_x$ ${(y \neq x)}$, where we have shown that $\{(\Omega-\tau)|_y = 0 , \xi|_y \neq 0\}$. We can follow the same steps as before, and since the condition on the matter fluid is global, we can conclude that $(\Omega-\tau)$ is vanishing and $\xi$ is non-vanishing in a neighborhood $\mathcal{N}_y$ of $y$ not entirely within $\mathcal{N}_x$. In this way, we can cover the entire manifold with overlapping open sets, each of which satisfies $(\Omega-\tau=0, \xi\neq0)$, thus proving that this statement is indeed valid globally. 
We can now define TLRS class III as follows:

\begin{definition}
TLRS class III  spacetimes are defined, in the comoving frame, by the property 
\begin{align}
\Omega-\tau=0 \ , \quad \xi\neq0  \ .
\end{align}
The manifold of this class only admits a foliation into a hypersurface orthogonal timelike vector field and a 3-D Riemannian submanifold. The matter fluid must satisfy the global conditions on the equation of state and hypermomentum
\begin{align}
    \mu+p(\mu)&\neq0 \ , & \tau=0 \ .
\end{align}
\end{definition}
Since torsion vanishes ($\tau=0$), this class is reduced to LRS class III in General Relativity. The matter fluid of a spacetime of this class can be a classical fluid satisfying $\mu+p>0$ (with the same physical interpretation as given in Sec. \ref{Classification-of-LRSGR-Subsection}) or a phantom matter fluid satisfying $\mu+p<0$ \citep{Dabrowski,LudwickPhantom}. The spacetimes belonging to TLRS class III further have the properties 
\begin{equation}
\phi=\mathcal{A}=\Omega=0\ ,
\end{equation}
 and $\hat{\psi}=0$ for any generic function $\psi$. Since torsion vanishes, the Weyl scalars are significantly simplified. In particular, from equations \eqref{TauHat}, \eqref{phihat} and (\ref{ETilde-Constraint}-\ref{HrbarHt-Constraint}), we obtain
\begin{equation}
\begin{gathered}
    \mathcal{H}_r = -\mathcal{H}_t = -\overline{\mathcal{H}}_t = \overline{\mathcal{H}}_r =3\xi\Sigma
    \ , \qquad
    \tilde{\mathcal{E}} = 0
    \ , \\
    \mathcal{E} = 2\xi^2 + \frac{2}{9}\Theta^2 + \frac{1}{3}\Sigma\Theta - \Sigma^2 - \frac{2}{3}\mu \ .
\end{gathered}
\end{equation}
Thus, the magnetic part $H_{ab} = \overline{H}_{ab}$ and the electric part $E_{ab}$ of the Weyl tensor are symmetric and traceless, as one would expect in General Relativity.

\subsection{TLRS Class IV}\label{sec-TLRSclassIV}
Let us now consider all the spacetimes which do not satisfy the hypersurface orthogonality conditions mentioned above, i.e., for which
\begin{align}
\Omega-\tau&\neq0 \ , & \xi&\neq0 \ , \label{C4condition1}
\end{align}
and therefore do not admit any foliation. These are certainly the most complicated cases to treat.  

An interesting case that can be associated with this type of spacetime is the one which allows
\begin{align}
    \mu+p&=0 \ , & \tau&=0 \ . \label{C4condition2}
\end{align}
With these assumptions, if we have a local statement on $\Omega-\tau$ and $\xi$ like \eqref{C1localvalue} or \eqref{C3localvalue}, then we have no way to control the behavior of the relevant kinematic quantities, as discussed in Sections \ref{sec-TLRSclassI} and \ref{sec-TLRSclassIII}. Thus, in general, in these manifolds, there might exist patches in which \eqref{C4condition1} is satisfied. In fact, the most general structure for TLRS class IV spacetimes can be imagined to consist of patches satisfying \eqref{C1localvalue} or \eqref{C3localvalue}, along with other patches in which \eqref{C4condition1} is satisfied. Since in these last patches, $\tau=0$, they are indistinguishable from a GR spacetime, and therefore, TLRS class IV spacetimes can also be seen as a mix between GR and ECSK theory.

We can then give the following definition:
\begin{definition}
TLRS class IV spacetimes are defined, in the comoving frame, by the property that
\begin{align}
\Omega-\tau\neq0 \ , \quad \xi\neq0  \ ,
\end{align}
or 
\begin{align}
    \mu+p=0 \ , \quad \tau=0 \ ,
\end{align}
is true (at a point, over a patch or globally). The manifolds of this class do not admit any foliation orthogonal to the congruences defining the comoving frame.
\end{definition}

From the discussion above, it should be evident that the definition of TLRS class IV and, most of all, of their properties is much weaker than the other classes.  Indeed, TLRS class IV can probably be better defined by simply stating that it is a class of locally rotationally symmetric spacetimes which do not belong to TLRS classes I, II, or III.

One can demonstrate that in TLRS class IV spacetimes, all scalars satisfy a second-order linear hyperbolic partial differential equation. In fact, the ${u^a e^b}$ projection of \eqref{DefofTorsion} gives the commutation relations for \textit{hat} and \textit{dot} derivatives
\begin{align}
    \dot{\hat{\psi}} - \hat{\dot{\psi}} &= \mathcal{A}\dot{\psi} - \left( \Sigma + \frac{\Theta}{3} \right)\hat{\psi} \ . \label{CommutationRelation}
\end{align}
Taking the \textit{hat} derivative of \eqref{DerCons}, the \textit{dot} derivative of \eqref{DerCons}, subtracting them and finally using \eqref{CommutationRelation} to eliminate the cross-derivative terms, we obtain the  partial differential equation which governs any scalar $\psi$ given as
\begin{align}
    \xi^2 \hat{\hat{\psi}} - (\Omega-\tau)^2 \ddot{\psi} &= 
    \xi (\Omega-\tau) (\Theta-3\Sigma)\hat{\psi} 
    \nonumber \\ & \quad
    + \xi (\Omega-\tau) (\mathcal{A}-2\phi) \dot{\psi}
     \ . \label{HyperbolicEqnofScalarIV}
\end{align}
However, the resolution of this equation is not immediate, and in addition, it is not necessarily true that the problem is mathematically well-posed. Indeed, we would need to prove the existence of a Cauchy surface to ensure that the covariant variables can be uniquely determined. However,  we know that every Cauchy surface is a (spacelike) embedded continuous submanifold of $\mathcal{M}$ (see e.g.  \citep{WaldBook}). As no foliation is possible for these spacetimes in the comoving frame, the existence of a global Cauchy surface remains doubtful.

\section{TLRS Class I Spacetimes}\label{Sec:ClassIB}
The spacetimes within TLRS class I do not expand ${(\Theta=0)}$ or distort ${(\Sigma=0)}$. As mentioned earlier in Section \ref{Classification-of-TLRSSpacetimes-subsection}, in these spacetimes the spacelike vector $e^a$ is non-twisting ${(\xi=0)}$. Further, the timelike vector field $u^a$, chosen to be the fluid-4 velocity, is a Killing vector field and, thus, these spacetimes are stationary (${\dot{\psi}=0}$, for a generic function $\psi$ on the manifold). However, $u^a$ is not hypersurface orthogonal ${(\Omega-\tau \neq 0)}$ , and therefore in this class it is not possible to define a global notion of time. The Weyl scalars, except $\mathcal{H}_t$, can be determined algebraically
\begin{equation}
\begin{aligned}
    \mathcal{E} &= -\mathcal{A}\phi + \frac{\mu}{3} + p - 2 \Omega^2
    \ , \\
    \mathcal{H}_r &= 2\mathcal{A}\tau - \Omega(2\mathcal{A} - \phi)
    \ , \\
    \overline{\mathcal{H}}_t &= \Omega(2\mathcal{A} - \phi)
    \ , \\
    \overline{\mathcal{H}}_r &= - \frac{1}{2} \left( \mathcal{H}_t + \overline{\mathcal{H}}_t \right)
    \ ,
\end{aligned}
\label{AlgebraicWeylTLRSIB}
\end{equation}
and we have ${\tilde{\mathcal{E}}=0}$. The propagation equations are
\begin{align}
    \hat{\Omega} - \hat{\tau} &= (\Omega - \tau)(\mathcal{A} - \phi)
    \ , \label{TLRSIBomegatau} \\
    \hat{\tau} &= -\tau\phi + \frac{\mathcal{H}_r}{2} + \frac{\mathcal{H}_t}{2}
    \ , \label{TLRSIBtau} \\
    \hat{p} &= -\mathcal{A} \left( \mu + p \right) -2\tau \overline{\mathcal{H}}_r
    \ , \label{TLRSIBp} \\
    \hat{\phi} &= -\frac{1}{2} \phi^2 + \mathcal{A}\phi + 2\Omega^2 - \mu - p 
    \ , \label{TLRSIBphi} \\
    \hat{\mathcal{A}} &= - \mathcal{A} \left( \mathcal{A}+\phi \right) - 2\Omega^2 + \frac{\mu}{2} + \frac{3}{2}p
    \ , \label{TLRSIBA}
\end{align}
and equation \eqref{ElecEnergyhat} is redundant. As before, in general, we need the equation of state ${p=p(\mu)}$ and the equation of spin density ${\tau=\tau(\mu)}$ to close the system of equations and to determine  $\mathcal{H}_t$. Indeed, one can also close the system by choosing the behavior of the magnetic Weyl scalars instead of imposing an equation of spin density on the matter fluid. This is not surprising since $\tau$ directly communicates with the conformal structure of the manifold through equations \eqref{BianchiINewEqns}; thus, it can be determined if the behavior of Weyl variables is known. As already mentioned, in this class, the matter fluid must satisfy at least one of the two conditions, either $\mu+p\neq0$ or $\tau\neq0$ at every point $x\in\mathcal{M}$.

In the following, it will also be useful  to combine  \eqref{TLRSIBomegatau} and  \eqref{TLRSIBtau} to obtain the equation
\begin{align}
    \hat{\Omega} &= \frac{1}{2} \left( \mathcal{H}_t -\Omega\phi \right) \ . \label{TLRSIBOmega}
\end{align}
We now focus on some specific examples of TLRS class I spacetimes.

\subsection{Gödel's Solutions with Torsion} \label{GodelUniverse}
We impose the condition ${\hat{\psi}=0}$ for any function $\psi$ on the manifold. Equation \eqref{TLRSIBomegatau} then gives (using $\Omega-\tau\neq0$)
\begin{align}
    \mathcal{A} &= \phi
    \ , \label{GodelRelation1}
\end{align}
Using  \eqref{GodelRelation1} in \eqref{TLRSIBtau} and \eqref{AlgebraicWeylTLRSIB}, we obtain an expression for the magnetic part of the Weyl tensor
\begin{equation}
\begin{aligned}
    \mathcal{H}_r &= 2\tau\phi - \Omega\phi
    \ , & 
    \overline{\mathcal{H}}_t &= \Omega\phi
    \ , \\
    \mathcal{H}_t &= 2\tau\phi - \mathcal{H}_r = \Omega\phi
    \ , &
    \overline{\mathcal{H}}_r &= -\Omega\phi
    \ .
\end{aligned}
\end{equation}
Instead, equations \eqref{TLRSIBphi} and \eqref{TLRSIBA}  can be used  to relate  the kinematic variables to the matter variables:
\begin{align}
    \phi^2 &= \frac{p}{3} - \frac{\mu}{3}
    \ , &
    \Omega^2 &= \frac{7}{12}\mu + \frac{5}{12}p 
    \ .
    \label{GodelOmegaphi}
\end{align}
Furthermore, the first equation of \eqref{AlgebraicWeylTLRSIB}  reduces to
\begin{align}
    \mathcal{E} = -\frac{\mu}{2} - \frac{p}{6} \ .
\end{align}
Thus, the only independent quantities remaining are ${\{ \mu,p,\tau \}}$, and all other quantities are determined algebraically. In particular, $\mathcal{H}_r$ requires information on torsion, while all the remaining quantities are determined by matter variables ${\{ \mu, p\}}$. 

Lastly, the substitution of the above conditions in equation \eqref{TLRSIBp} gives us
\begin{align}
    \phi (\mu + p - 2\tau\Omega) = 0 \ . \label{Godelcases}
\end{align}
Thus, we can discuss two possible cases: 
\begin{enumerate}
\item \label{Godelphizerocase} Setting ${\phi=0}$, we obtain
\begin{gather}
    \mathcal{A} = \mathcal{H}_r = \mathcal{H}_t = \overline{\mathcal{H}}_r = \overline{\mathcal{H}}_t = 0
    \ , \\
    \Omega^2 = p = \mu= -\frac{3}{2}\mathcal{E}  \ .
\end{gather}
Given $\mu$, the rest of the variables, but $\tau$, can be algebraically determined. Notice that the torsion variable $\tau$ is left independent, i.e., it does not influence the kinematics of the spacetime. In this case, all the covariant variables are the same as one would have had in the Gödel's universe of General Relativity. However, torsion still affects the behavior of the timelike congruences. Based on the choice of $\mu$ and $\tau$, this spacetime can belong to TLRS class IA or IB or both.

\item For ${\phi \neq 0}$, Eq. \eqref{Godelcases} implies 
\begin{equation}
\mu+p=2\tau\Omega \ .
\end{equation} 
Squaring this condition and using \eqref{GodelOmegaphi}, we obtain
\begin{align}
    \tau^2 = 3\frac{(\mu+p)^2}{7\mu+5p} \ . \label{tauofgodel2}
\end{align}
Then, given an equation of state and the energy density, one can determine all other covariant variables. As the conditions $\mu+p\neq0$ and $\tau\neq0$ are simultaneously satisfied, this spacetime belongs to both TLRS class IA and IB. Additionally, the matter fluid must satisfy
\begin{align}
\qquad \quad
    p > -\frac{7}{5} \mu \ ,
\end{align}
to avoid divergence $\tau \rightarrow \infty$ and $\tau^2<0$. It is useful to look at these conditions in terms of projections of the metric energy-momentum tensor $s_{ab}$. Using  ${(\mu+p=2\tau\Omega)}$ and equation \eqref{metric-canonical-LRSrelations}, one can evaluate
\begin{equation}
\begin{aligned}
    \qquad \quad
    \overline{\mu} &= -\mu -2p
    \ , &
    \overline{p} &= -\frac{2\mu}{3} + \frac{p}{3}
    \ , \\
    \qquad \quad
    \overline{\Pi} &= \frac{2}{3}(\mu+p)
    \ , & 
    \overline{q} &= 0
    \ ,
\end{aligned}
\end{equation}
which also leads to the equation of state
\begin{align}
    \overline{\mu} + \overline{p} + \frac{5}{2}\overline{\Pi} = 0
    \ . \label{EquationofStateGodel2}
\end{align}
Defining  the \textit{radial pressure} and the \textit{shearing pressure} as
\begin{equation}
\overline{p}_r = \overline{p} + \overline{\Pi} \ ,\qquad \overline{p}_t = \overline{p} - \frac{1}{2} \overline{\Pi} \ , \label{radialtangentialpressure}
\end{equation}
equation \eqref{tauofgodel2} can  be written as
\begin{align}
    \tau^2 = \frac{3(\overline{\mu} + \overline{p}_r)^2}{5\overline{p}_r - 7\overline{p}_t} \ .\label{tauofgodel2metricenergy}
\end{align}
Thus, to ensure the condition $\tau^2>0$ is satisfied, we require ${\overline{\mu}+\overline{p}_r \neq 0}$ and ${5\overline{p}_r > 7\overline{p}_t}$.
\end{enumerate}

\subsection{Torsional Spacetimes Without The Magnetic Part Of The Weyl Tensor}\label{SubSec:VoritcalNoWeylST} 
We now consider TLRS  spacetimes of class I with the condition 
\begin{align}
    \mathcal{H}_r = \mathcal{H}_t = \overline{\mathcal{H}}_r = \overline{\mathcal{H}}_t=0 \ ,
    \qquad
    \tau\neq0 \ . \label{nomagcondition}
\end{align}
i.e., a gravitationally silent torsional spacetime, in which there are no gravitational waves. These manifolds can be divided into two subcases, ${\Omega \neq 0}$ and ${\Omega = 0}$. This is because equation \eqref{TLRSIBOmega} now reduces to
\begin{align}
    \hat{\Omega} = -\frac{1}{2} \phi \Omega \ ,
\end{align}
and therefore, if $\Omega=0$, then it remains zero since the above equation leads to an exponential solution for $\Omega$ in a small enough neighborhood. 

For ${\Omega \neq 0}$, we get back Gödel's solution with torsion for ${\phi=0}$ in Section \ref{GodelUniverse}. 

For ${\Omega = 0}$, instead, \eqref{AlgebraicWeylTLRSIB} and \eqref{nomagcondition} give
\begin{align}
    \mathcal{A} = 0 \ , \label{NoGWrelation1}
\end{align}
which along with \eqref{TLRSIBA} leads to
\begin{align}
p &= - \frac{\mu}{3} \ . \label{EOSofCurvature}
\end{align}
Using the results above, the electric part of the Weyl tensor also vanishes (using  \eqref{AlgebraicWeylTLRSIB})
\begin{align}
\mathcal{E} = 0 \ .
\end{align}

Hence, the entire Weyl tensor vanishes, the matter follows the geodesic path (${\mathcal{A}=0}$), there is no vorticity (${\Omega=0}$), and the matter fluid has an equation of state given by \eqref{EOSofCurvature}. The propagation equations \eqref{TLRSIBtau}, \eqref{TLRSIBp} and \eqref{TLRSIBphi} are simplified to
\begin{align}
    \hat{\tau} &= -\tau\phi
    \ , &
    \hat{p} &= 0
    \ , &
    \hat{\phi} &= -\frac{\phi^2}{2} + 2p
    \ . 
    \label{NoWeylCovariantEqn}
\end{align}
Thus, the matter variables $\{ \mu=\mu_0, p=p_0\}$ are constant in spacetime. Since we have used \eqref{TLRSIBOmega} (which vanishes for this subcase), \eqref{TLRSIBomegatau} is now redundant. Further, since $\Omega=0$, we obtain from \eqref{metric-canonical-LRSrelations} that projections of the metric energy-momentum tensor are given as
\begin{align}
    \overline{\mu} = \mu_0
    \ , \quad
    \overline{p} = p_0
    \quad \implies \quad
    \overline{p} = -\frac{\overline{\mu}}{3}
    \ ,
\end{align}
while we have $\overline{\Pi}=0$ and $\overline{q}=0$. We have then obtained a gravitationally silent torsional spacetime dominated by dark radiation.

Equations \eqref{NoWeylCovariantEqn} can be solved exactly. Choosing a spacelike coordinate function $r$ such that $e^a$ is given as\footnote{The choice of coordinate is based on ease of evaluation. Coincidently, this coordinate function is similar to the area-radius coordinate. Therefore, one can write the interval of the 2-surface described by $N_{ab}$ as ${dl^2{}_{2D} = r^2 (d\theta^2 + \sin^2{\theta} d\varphi^2)}$, for angular coordinates $\theta$ and $\varphi$. However, since $u^a$ is not hypersurface orthogonal, we do not have a $2-$dimensional integral submanifold, and $u^a$ can also contribute to $d\theta^2$ and $d\varphi^2$ terms in the spacetime interval of the manifold $\mathcal{M}$.}
\begin{align}
    e_a &= \frac{2}{r\phi} dr 
    \ , & 
    e^a &= \frac{r\phi}{2}\frac{\partial}{\partial r}
    \ ,
\end{align}
the solution of \eqref{NoWeylCovariantEqn} can be evaluated to be
\begin{align}
    \tau &= \frac{\tau_0}{r^2}
    \ , &
    \phi^2 &= 4p_0 + \frac{2c}{r^2}
    \ ,
\end{align}
where $c$ and $\tau_0$ are constants of integration. This spacetime belongs to TLRS class IB by definition and also belongs to TLRS class IA if one chooses $p_0\neq0$.

\subsection{Spacetimes in Canonical Vacuum}\label{TLRS1B-CanonicalVacuum}
Let us now consider geometries generated by canonical vacuum, $S_{ab}=0$, described in Section \ref{Sec:1+1+2Decomposition}. Since we violate $\mu+p\neq0$, we must impose $\tau\neq0$, and the spacetime belongs to TLRS class IB only. From \eqref{TLRSIBp}, we have
\begin{align}
    \overline{\mathcal{H}}_r = 0 \ .
\end{align}
 Using \eqref{AlgebraicWeylTLRSIB}, the rest of the Weyl variables can be determined algebraically
\begin{equation}
\begin{aligned}
    \mathcal{E} &= -\mathcal{A}\phi - 2 \Omega^2
    \ , &
    \mathcal{H}_r &= 2\mathcal{A}\tau - \Omega(2\mathcal{A} - \phi)
    \ , \\
    \overline{\mathcal{H}}_t &= \Omega(2\mathcal{A} - \phi)
    \ , &
    \mathcal{H}_t &= -\Omega(2\mathcal{A} - \phi)
    \ ,
\end{aligned}
\end{equation}
and the governing propagation equations are (simplified from  \eqref{TLRSIBomegatau}, \eqref{TLRSIBphi}, \eqref{TLRSIBA}, \eqref{TLRSIBOmega})
\begin{align}
    \hat{\Omega} - \hat{\tau} &= (\Omega-\tau)(\mathcal{A}-\phi)
    \ , \\
    \hat{\phi} &= -\frac{\phi^2}{2} + \mathcal{A}\phi + 2\Omega^2
    \ , \\
    \hat{\mathcal{A}} &= -\mathcal{A}(\mathcal{A}+\phi) - 2\Omega^2
    \ , \\
    \hat{\Omega} &= -\mathcal{A}\Omega \ .
\end{align}
Using \eqref{ProjMetricEMCanVacuum}, we have the following relations between projections of the metric energy-momentum tensor
\begin{align}
    \overline{p} &= \frac{\overline{\mu}}{3}
    \ , &
    \overline{\Pi} &= -\frac{\overline{\mu}}{3}
    \ ,
\end{align}
with $\overline{q}=0$. These relations can be presented in terms of radial pressure $\overline{p}_r$ and shearing pressure $\overline{p}_t$ (defined in equation \eqref{radialtangentialpressure}) as
\begin{align}
    \overline{p}_r &= 0
    \ , &
    \overline{p}_t &= \frac{\overline{\mu}}{2}
    \ .
\end{align}
Thus, the matter source only has shearing pressure and no radial pressure. Therefore, this spacetime represents a stationary gravitational field generated by a distribution of matter for which only shearing pressure appears. This configuration is analogous to the one found by Florides in LRS-II spacetimes \cite{Florides1974}  and raises the question of the existence, within ECSK theory, of vortical compact stars.

\section{Discussion and Conclusions}\label{DiscussConclusion}
In this paper, we have derived the complete set of covariant equations that describe spacetimes possessing local rotational symmetry and non-vanishing torsion and are sourced by the Weyssenhoff spin fluid in ECSK theory. The covariant approach we employed allows us to assign distinct physical meaning to the covariant variables that characterize spacetime, making it much easier to study its physical properties. Since this approach prevents confinement to a specific coordinate system, one avoids making coordinate (gauge) dependent conclusions (for example, coordinate singularity). Both these properties make this framework particularly suitable for any further investigation into locally rotationally symmetric spacetime with Weyssenhoff-like torsion.

We found that in the ECSK theory, the Weyl tensor becomes much more complicated.  Such richness in the conformal structure of the manifold directly relates to the presence of torsion, as can be seen from the projections of the type-I Bianchi identity. One of the most significant distinctions between non-torsional and torsional locally rotationally symmetric spacetimes is that the magnetic part of the Weyl tensor is no longer algebraically characterized by kinematic variables only. Indeed, the covariant scalar $\mathcal{H}_t$ depends on the covariant derivative of the torsion parameter $\tau$ along $e_a$. This result indicates that the structure of gravitational waves on TLRS spacetimes can be substantially different from that of their LRS counterparts. Notice also that one can omit the need for the equation of spin density to close the system of equations if one imposes a specific conformal structure of the manifold, as shown in the example in Section \ref{SubSec:VoritcalNoWeylST}.

We also presented the condition such that the manifold admits a foliation into a hypersurface orthogonal one-form and an integral $3-$dimensional submanifold. Understanding the hypersurface orthogonality condition shows that, differently from GR,  it is more appropriate to look at the behavior of the combination ${(\Omega-\tau)}$ rather than $\Omega$. In addition, a precise definition of hypersurface orthogonality allows us to appreciate the relationship between the classification of the LRS spacetime in General Relativity and the foliation admitted by the spacetime manifold for each class.

Furthermore, we have proposed a classification scheme for TLRS spacetimes based on the foliation admitted by the manifold and showed that such a classification scheme can be successfully applied. In particular, we found four classes of spacetime, dubbed TLRS class I, II, III, and IV, which, similarly to the GR case, are distinguished by the admittance of a foliation into spacelike or timelike hypersurfaces and some conditions on torsions and thermodynamics. The first three classes are separated in the sense that the kinematic or thermodynamic properties that define them are global and, therefore, only valid in that class. This was shown by taking a local value (vanishing or non-vanishing) of defining variables of the TLRS classes, namely $\Omega-\tau$ and $\xi$, and demonstrating that those values remain true over the entire manifold. Thus, a spacetime having either $\Omega-\tau$ or $\xi$ (or both) zero locally will present the same properties also globally.  

In TLRS class I spacetimes, the matter fluid needs to satisfy at least one of the conditions, $\mu+p\neq0$, or $\tau\neq0$ at every point on the manifold. While $\mu+p\neq0$ is a generalization of the condition necessary for classification in General Relativity,  the condition $\tau\neq0$  shows that this class contains spacetimes which are unique to ECSK gravity sourced by a Weyssenhoff fluid. Within the TLRS class I, we can distinguish two subclasses, IA and IB, characterized by either $\mu+p\neq 0$ or $\tau\neq 0$. Indeed, one can even have TLRS class I spacetimes in which the condition $\mu+p\neq0$ is realized in some part of the manifold and $\tau\neq0$ on the complementary part of the manifold. Such a configuration could find applications in modeling relativistic stellar solutions with non-vanishing torsion.

For TLRS class II, the matter fluid does not need to satisfy any condition. We can contrast this with classification in General Relativity and LRS class II, in which one always pre-supposes a condition on the matter fluid. Since our classification scheme is primarily based on the foliation admitted by the manifold, we were able to omit any unnecessary condition on the matter fluid. Notice that our approach could be used in GR to extend the traditional classification of LRS spacetimes to cases not traditionally included (see  \citep{CarloniTOVAniso1Fluid,CarloniNaidu} for some examples).

Unlike the previous two classes, we found that TLRS class III is always torsion-free. This result implies that TLRS class III in ECSK gravity with Weyssehoff fluid contains the same spacetimes as LRS class III in General Relativity.  We believe that this is a result of the special form of Weyssenhoff torsion. Indeed, the ultimate reason why TLRS class I presents differences with respect to LRS class I is that the hypersurface orthogonality of $u^a$ is characterized by the quantity $\Omega-\tau$ rather than just $\Omega$, and this is essentially due to \eqref{SimplifiedTorsionofTheory}. For example, should a term containing $e^c\eta_{ab}$ appear in that relation, we could not use $\xi$ to characterize hypersurface orthogonality, and there would be space for a richer TLRS class III.

We have collected the TLRS spacetimes that do not belong to any of the TLRS classes above in TLRS class IV. This class differs from the other three above in that it either derives its meaning from the matter variables as spacetimes with existence of regions with $\mu+p=0$ and $\tau=0$, or otherwise, it is counter-intuitively defined via a lack of a global foliation orthogonal to the congruences describing the fluid rest frame. Indeed, TLRS class IV spacetimes can even be formed by patches which mimic the foliation of TLRS class I and III and patches with $\Omega-\tau\neq0$, $\xi\neq0$. These last regions can be thought of as buffers that connect the local TLRS class I  and  III regions (if any are present).

The absence of a global foliation for the spacetime complicates the treatment of TLRS IV spacetimes. Indeed, one can prove that a second-order linear hyperbolic partial differential equation exists for the covariant variables. However, conclusive proof of the applicability of the Cauchy theorem is not immediately obtainable. This means that this kind of spacetime should most likely be treated locally.  Despite these complications, TLRS Class IV remains the most physically realistic among the four classes. This becomes evident if we think that in the observed Universe, there is vorticity and twist, often at the same time, and that the presence of non-adiabatic processes means that a flux term is always present.

The above classification has been constructed by choosing to describe spacetimes in a frame that is comoving with the Weyssenhoff fluid. Thus, the definition and the properties of spacetimes of each class are frame specific. In principle, a classification can be formulated in a {\it generic} frame, as it is done in GR.
In GR, a global matter condition $\mu+p\neq0$ leads to existence of only three classes as discussed. Therefore, LRS class IV would be characterized by the violation of this thermodynamic condition (instead of a statement on its foliation) and probably for this reason, this class is rarely introduced.
This invokes the question if TLRS class IV would also be more suitably defined via thermodynamic conditions (for classification formulated in a generic frame).

As an application of the 1+1+2 equations and the TLRS classification, we studied various examples of TLRS class I. In Section \ref{GodelUniverse}, we discussed a Gödel-like solution with torsion, which has the same properties as the Gödel Universe in General Relativity, except it allows for a non-vanishing independent torsion. One can, therefore, expect that the presence of torsion will affect the geodesics of timelike congruences and the formation of closed time loops, opening the way to interesting scenarios. In Section \ref{SubSec:VoritcalNoWeylST},  instead, we presented an example showing that imposing a conformal structure on the manifold omits the necessity for information on an equation of state and an equation of spin density to close the 1+1+2 equations.  More specifically, we found a solution representing a gravitationally silent universe dominated by dark radiation. By breaking covariance, we have been able to resolve the covariant equations exactly in terms of a parameter similar to the area radius. As a final example, in Section \ref{TLRS1B-CanonicalVacuum}, we studied TLRS I spacetimes in canonical vacuum. While this system appears to be unphysical for TLRS classes II and III because it leads to a negative energy density $\overline{\mu}$, it can yield physical results in TLRS class I, particularly for subclass IB. Interestingly, this example yields a solution in which matter is supported solely by a shearing pressure. It is known that the Florides solution, when interpreted as a stellar object, requires a shell to be stable, but this might not be the case within its counterpart in ECSK theory.  

It is also worth stressing that since TLRS class I spacetimes do not admit foliation orthogonal to a timelike congruence, a notion of global time (like cosmic time) would be absent from any further physical interpretation/application of such examples. On the other hand, the fact that these solutions are stationary ($\Theta=0=\Sigma$) makes this shortcoming less relevant in the context of the potential physical value of these solutions.

As discussed, the results of this paper form a framework that can be utilized to study torsion spacetimes with local rotational symmetry and their physical properties and develop a taxonomy of exact solutions. While in our case, the classification was only done in the comoving frame for Weyssenhoff-like torsion, we have no reason to doubt that the scheme described in Sec.\ref{Classification-of-TLRSSpacetimes-subsection} would also be successful for any type of torsion and/or source fluid or in a generic frame. At the same time, the complete set of equations can be utilized to extend the understanding of other aspects of ECSK theory. For example, they can be employed to formulate the junction conditions in the ${1+1+2}$ covariant formalism, which would extend the work in Ref. \citep{RosaCarloni} to torsion spacetimes. Another more complex application would be in the perturbative study of the static solutions of the interior of stellar objects as discussed in \citep{CarloniLuzGeneralPert,SanteLuzPertTilt,SanteLuzPertComoving} in the case of GR.

\appendix

\section{Integrable Submanifolds}\label{Appen-IntegrableSubmanifold}
Let $\mathcal{M}$ be a manifold of dimension${-d}$, ${T\mathcal{M}}$ be the tangent space on the manifold, and ${T^*\mathcal{M}}$ be the co-tangent space on the manifold. 

Let $U$ be a collection of vector fields of dimension${-n}$ ($n<d$) such that for any point ${p \in \mathcal{M}}$, $U_p \subset {T_p\mathcal{M}}$, where ${T_p\mathcal{M}}$ is the tangent space at point ${p \in \mathcal{M}}$, and the collection can be generated by $n$ smooth vector fields which are members of $U$. Then $U$ forms a subspace of the tangent space $T\mathcal{M}$.

Similarly, let $U^*$ be a collection of co-vector fields of dimension${-m}$ ($m<d$) such that for any point ${p \in \mathcal{M}}$, $U^*_p \subset {T_p^*\mathcal{M}}$, where ${T_p^*\mathcal{M}}$ is the co-tangent space at point ${p \in \mathcal{M}}$ and the collection can be generated by $m$ smooth co-vector fields which are members of $U^*$. Then $U^*$ forms a subspace of the co-tangent space $T^*\mathcal{M}$.

The subspaces $U$ and $U^*$ are complementary if ${m+n=d}$ and we have 
\begin{equation}
X^a \alpha_a = 0, \quad \forall X \in U, \ \forall \alpha \in U^* \ .  
\end{equation} 

A tangent subspace $U \subset T\mathcal{M}$ is said to be involutive if we have 
\begin{equation}
 [X,Y]\in U , \quad \forall X,Y \in U \ ,  
\end{equation} 
where the \textit{square brackets} ${[\cdot{},\cdot{}]}$ denotes Lie bracket. In other words, $U$ is involutive if
\begin{equation}
\alpha_a [X,Y]^a = 0, \quad \forall X,Y \in U\ ,  
\end{equation}
where $\alpha$ is a member of the complementary co-tangent subspace.

The cotangent subspace $U^* \subset T^*\mathcal{M}$ is said to be differential if we have
\begin{equation}
X^a Y^b D\alpha_{ab} = 0 , \quad {\forall \alpha\in U^*}\ ,  
\end{equation}
where $X,Y$ are members of complementary tangent subspace, and $D(\cdot)$ is the exterior derivative. In other words, for the generators of $U^*$, given as $\alpha^\mu$, we must have (written without abstract indices/components)
\begin{equation}
D\alpha^\mu = \Theta^\mu_\nu \wedge \alpha^\nu\ ,  
\end{equation}
where $\Theta^\mu_\nu\in {T^*\mathcal{M}}$ is a one-form, and $\mu,\nu=1,...m$ index the generators $\alpha^\mu$ and $\Theta^\mu_\nu$. 

\bibliography{biblio}
\end{document}